\RequirePackage{fix-cm}
\documentclass[smallextended]{svjour3}       
\smartqed  
\usepackage{lscape}
\usepackage{natbib}
\usepackage{graphicx}
\usepackage{mathptmx}
\usepackage{amsmath}
\usepackage{latexsym}
\usepackage{graphicx}
\usepackage{IEEEtrantools}
\begin{document}

\title{Is it Possible to Disregard Obsolete Requirements?
}
\subtitle{A Family of Experiments in Software Effort Estimation}
\author{
}
\institute{
}

\author{Lucas Gren         \and
        Richard Berntsson Svensson 
}
\institute{L. Gren \at
              Blekinge Institute of Technology \\
               Karlskrona, Sweden\\
Volvo Cars and Chalmers $|$ University of Gothenburg \\
Gothenburg, Sweden\\
              \email{lucas.gren@bth.se}           
           \and
           R. Berntsson Svensson \at
                          Chalmers University of Technology and The University of Gothenburg \\
Gothenburg, Sweden\\
              \email{richard@cse.gu.se}   
}

\date{Preprint accepted for publication in Requirements Engineering Journal March 18, 2021}

\maketitle

\begin{abstract}
\emph{Context}
Expert judgement is a common method for software effort estimations in practice today. Estimators are often shown extra obsolete requirements together with the real ones to be implemented. Only one previous study has been conducted on if such practices bias the estimations. 

\emph{Objective}
We conducted six experiments with both students and practitioners to study, and quantify, the effects of obsolete requirements on software estimation. 

\emph{Method}
By conducting a family of six experiments using both students and practitioners as research subjects ($N=461$), and by using a Bayesian Data Analysis approach, we investigated different aspects of this effect. We also argue for, and show an example of, how we by using a Bayesian approach can be more confident in our results and enable further studies with small sample sizes. 

\emph{Results}
We found that the presence of obsolete requirements triggered an overestimation in effort across all experiments. The effect, however, was smaller in a field setting compared to using students as subjects. Still, the over-estimations triggered by the obsolete requirements were systematically around twice the percentage of the included obsolete ones, but with a large 95\% credible interval.

\emph{Conclusions}
The results have implications for both research and practice in that the found systematic error should be accounted for in both studies on software estimation and, maybe more importantly, in estimation practices to avoid over-estimations due to this systematic error. We partly explain this error to be stemming from the cognitive bias of anchoring-and-adjustment, i.e.\ the obsolete requirements anchored a much larger software. However, further studies are needed in order to accurately predict this effect. 

\end{abstract}

\section{Introduction}\label{intro}
In all types of project, the planning phase includes some kind of effort forecasting. Since the 1940s, researchers have been investigating the use of expert opinion in connection to getting as accurate estimations as possible \citep{delphi}. Many aspects have been studied in relation to software cost estimation due to an explosion of software-related projects in the last decades \citep{halkjelsvik2012from}. Many of these studies have empirically investigated the impact of irrelevant information (i.e.\ information that is not needed for the estimations) on software effort estimations. In \citet{jorgen2}, the results show that pre-planning effort estimates may have an impact on the detailed planning effort estimates, despite subjects being told that the early estimates are not based on historical data. Furthermore, \citet{jorgen3} report that, despite that the subjects were told that customer expectation is not an indicator of the actual effort\footnote{We use the terms ``effort,'' ``cost,'' and ``time'' interchangeably when discussing estimation in this paper because the main driver for cost is typically the effort in connection to software development, which takes time from employees that is payed for by the organizations.}, irrelevant information about the customer's expectations affects the cost estimates. In addition, the results in \citet{jorgensen} indicated that the length of the Requirements Specification had an impact, however small, on the effort estimates. Finally, in a study by \citet{aranda}, the results show that information that is clearly marked as irrelevant (i.e.\ not to be taken into account) in a requirement specification have a large impact on software cost estimates. The results in \citet{aranda} could not be explained by the subjects' experience of cost estimations. \citet{aranda} explicitly tested the cognitive bias of \emph{anchoring} and concluded that an estimate from a clearly-stated non-expert was still influencing the judgement of the participants. In general, the above mentioned studies have shown that introducing irrelevant information may lead to an increased estimation error, but with a small sample sizes of around 20 participants in each study, which implies low statistical power.

One aspect that has not been studied, except for an initial study \citep{gren2017possible}, is the effect of obsolete requirements, i.e.\ requirements that are somehow marked as not to be included in the estimations but still visible to the assessors when estimating. The reason why this aspect should be studied more, is that the way software development practice often deals with requirements that should not be implemented now is to mark them as ``obsolete'' or the like \citep{wnuk2013obsolete}, which is a special type of irrelevant information. In our experience, most companies have too many requirements and it is not possible to implement all of them in the coming product\slash project\slash release, or in the next sprint for companies using an agile software development processes. It is, of course, then important to make accurate estimates of the ones that actually are to be implemented. If current approaches misguide the effort estimation, the practices must of course change or at least be informed by the impact of showing obsolete requirements to estimators.

\subsection{Previous research and motivation}\label{sec:ade}
The first study conducted on the topic of obsolete requirements was published in \citet{gren2017possible}. In order to clarify the experimental setup (more details are available in \citet{gren2017possible}) the authors distributed three different tasks to three groups of students in the same class. The first group (group A) was to estimate how long time it took to implement 4 requirements in weeks. Group B was given the same 4 requirements plus one extra (a total of 5 requirements). Group C was given the same 5 requirements as Group B but was instructed to leave the last requirement out of the estimation. 

The study was conducted with 150 university students and showed that adding obsolete requirements to the requirement specification heavily distorted the students and manipulated them into providing higher estimates for the existing requirements (i.e.\ they provided much lower estimates without any requirements marked as obsolete next to them). Before the experiment started, a pre-questionnaire was given to the students to collect the students experiences and knowledge in relation to the English language, experience from software development in industry, and experience in effort estimation. The study was conducted during one lecture in the mandatory course. 

In total, the experiment lasted for one hour, including introduction, explanations, pre-questionnaire, and completing the tasks. The actual time spent on the tasks, including read the instructions and performing the estimation, was 10-20 minutes. The task groups, i.e.\ A, B, and C were not overlapping. That is, the 150 students were divided into three groups for the three different estimation tasks. Meaning, 50 students performed task A (was in Group A), 50 students performed task B (Group B), and 50 performed task C (Group C). A more detailed description of the experiment, the experimental material, subjects, and setup is available in \citet{gren2017possible}.

Intuitively, if the same specification is used but some additional requirements are market as obsolete in one group, these estimates should be similar, and preferably be estimated as if those requirements were not there since they were explicitly marked as obsolete. However, the result showed that the estimates instead increased heavily. The authors tried to explain the effect by suggesting that two different cognitive biases could play a role, namely the representativeness heuristic \citep{representativeness} or the decoy effect \citep{zang}. However, none of these explanations helped in quantifying the effect of obsolete requirements, which is why we decided to investigate how the found estimation bias functions in more detail by conducting further experiments. 

\subsection{Research goal and research question}
The aim of this current family of experiments is to investigate the effects of obsolete requirement in software effort estimation further through a set of six experiments. The experiments comprise of both student participants estimating real requirements individually (Experiments 1, 2, and 3), industry practitioners doing the same (Experiment 4), and industry practitioners estimating their own requirements (Experiment 5), and the same industry practitioners as in Experiment 5 estimating their own requirements in teams over time in sprints (Experiment 6). Therefore, the overall research question we looked at from different angles is:

\emph{RQ: Do obsolete requirements, explicitly stated or marked to be excluded from the effort estimation, have an impact of the size of the total estimates? and if so, how much?}

\subsection{Contribution}
This paper contributes with a family of six experiments to show the effect of obsolete requirements in different context and with different requirements specifications, which was large across all experiments. Moreover, this paper shows how Bayesian Data Analysis (BDA) can be used to statistically analyze studies without the use of statistical significance. By using BDA, this paper enables replications with very small sample sizes since new experiments can use what have already been learned about the parameters in this study.

The remaining paper is organized as follows. In the next section (Section~\ref{bayes}), we provide a brief introduction to Bayesian Data Analysis (BDA). Section~\ref{family} presents an overview the six experiments conducted in this current study and if\slash how we changed the experimental setup after each experiment. In Section~\ref{sec:results} we show the output from each experiment. In Section~\ref{sec:disc} we discuss the findings from all the experiments, in Section~\ref{sec:limits} we discuss threats to validity, and in Section~\ref{sec:concl}, we conclude the paper and suggest future work.

\section{Bayesian Data Analysis}\label{bayes}
We have lately followed the development in statistics with great interest (e.g.\ \citet{munafo2017manifesto}), but a great summary that inspired the data analysis used in this study is the recent publication by \citet{mcshane2019abandon} where they argue for researchers to abandon statistical significance completely. Their remedy is the use of something that can be denoted a ``fully'' Bayesian Data Analysis (BDA) with no threshold values but an open and honest presentation of prior beliefs, data, and all the analyses conducted. In 2019, a first paper was published in software engineering critiquing current statistical practice, and suggesting BDA as a potential solution \citep{furia2019bayesian}. 

Any statistical investigation have data from a random variable from a probability distribution $P$. In most software engineering research, this distribution if often assumed to be normal (i.e.\ Gaussian), and if not, assumed to not exist and instead researchers use statistical tests based on ranks \citep{de2019evolution}. However, this is a pity since there are many probability distributions that could create much better models for the collected data (e.g.\ Binomial, Beta, Poisson, Log-normal, etc.). All these distributions can be described by parameters, $\theta$s. When researchers have conducted a study, some data $D$ is collected, but assumptions need to be made, or preferably, trying to find the best fitting distribution for the data. It is important to stress here that any statistical inference eventually make use of Bayes' theorem \citep{McElreath2016sra}, but a Bayesian Data Analysis approach uses this theorem more generally and in connection to parameters and models. Bayes' theorem yields:

\begin{equation}
P(\theta \mid D) = \frac{P(D | \theta) \times P(\theta)}{P(D)}
\end{equation}
Where $P(\theta \mid D)$ is the probability of the parameter $\theta$ given the data. This is called the \textit{posterior distribution} and is what should be obtained in the end for all the parameters of interest. Once the posterior is obtained, it is possible to analyze it from different perspectives and make inferences. $P(D | \theta)$ is the \textit{likelihood} that the data actually came from the assumed parameter. It is important to try different likelihoods, i.e.\  statistical models including different statistical distributions for each parameter, and compare how these different scenarios affect the posterior. $P(\theta)$ is the \textit{prior} information about the parameter, which is then not connected to the obtained data. $P(D)$ is simply a standardizing constant, expressed as the average likelihood.

It is rarely possible to exactly calculate the posterior distribution, which is why we instead sample from the posterior using Markov-Chain Monte-Carlo simulation. This is one of the reasons BDA was less used before modern computers with enough computing power for such sampling methods \citep{McElreath2016sra}. 

As mentioned, BDA is not about Bayes' theorem, but about quantifying uncertainty much more than the frequentist approach. We can try different likelihoods, use the prior information about parameters and integrate all these into a model that include all the uncertainty for all the parameters. An controversy in BDA is the choice of priors since they will affect the results to a very large extent. Therefore, one uses weakly informative priors if no prior information exist and then uses the posterior from earlier studies in the future. What should also be done, since practical significant is the ultimate goal in research, is to use experts to provide this prior information \citep{stefan2019practical}. For a short background of BDA and why it is useful for software engineering research, we refer to \citet{furia2019bayesian}. For an example of a good text from another research field, see \citet{van2014gentle}. 

We would recommend readers interested in learning BDA to first read the book by \citet{McElreath2016sra} and try our the R package Rethinking\footnote{https://github.com/rmcelreath/rethinking}, and then go from defining models in Rethinking to brms \citep{burkner2017brms}, which is faster and simpler to do more advanced analyses, but less pedagogical. Both packages build on R\footnote{https://www.r-project.org/} and Stan\footnote{https://mc-stan.org/}.


Other researchers lead the development of BDA and we will only apply it in this paper. We followed the steps below, which can be read about in much more detail in \citet{wilson2019ten} (some of which can be followed in the Supplementary Material):
\begin{enumerate}
    \item Always plot the raw data to get an initial idea of what the distributions might be for what we have collected. 
    \item Create an initial statistical model and check how it behaves without looking at the data (i.e.\ a sensitivity analysis). 
    \item Create different models and obtain posterior distribution for all of them (i.e.\ the models in light of the data) and validate them against each other.
    \item Check how the chains behave in the Markov-Chain Monte-Carlo simulations to find the posteriors.
    \item Plot and look at the real distributions of the posteriors to assess the results.
    \item Calculate a Bayesian $R^2$ statistic \citep{gelman2019r} to assess variance explained by the model, but by using the posterior. 
\end{enumerate}

\section{A Family of Experiments}\label{family}
In order to investigate the estimation bias, we conducted six experiments (in addition to the experiment conducted by \citet{gren2017possible} from whom we obtained the raw data) with both students and practitioners ($N=461$) to see if obsolete requirements explicitly stated to be excluded from the effort estimate had an impact on the size of these estimates.

Hereinafter, we denote the experiment published in \citet{gren2017possible} as Experiment 0 since it was the first one to be conducted on this topic but not a part of this current paper. Assessing the validity threats of Experiment 0 \citep{gren2017possible}, there is an evident problem with instructing subjects to exclude requirements on their paper next to the requirements, which is why we replicated the experiment in a set of different settings in this paper. In more detail, it may be confusing to read the phrase ``Requirement x should not be implemented,'' which is why the experiment was replicated in an as realistic setting as possible (Experiment 6). Regarding Experiment 0, first, it is not known if the results from Experiment 0 replicates with exactly the same setup (addressed in Experiments 1 and 2). Second, it was not possible to know if the length of the requirement specification is a confounding factor (addressed in Experiments 3 and 4) or if the effect might disappear by conducting the estimation in teams (addressed in Experiments 6), which many companies do. Also, having students estimate requirements (Experiments 1, 2, and 3) they know they will not implement for a system they are unfamiliar with has, of course, a great risk of being a toy problem. Experiment 4, therefore, comprised of industry participants, but they still estimated requirements they were not to implement by themselves afterward. Experiments 5 and 6 looked at this aspect by being fully in context of developers that both estimated and later implemented the requirements. 

Furthermore, the first set of experiments (Experiments 0--3) did not investigate the accuracy of the estimates since we did not compare to an actual implementation effort (we did not obtain ``true'' student implementation times). It could have been the case that the obsolete requirements helped the subject to decrease the estimation error. Therefore, in Experiment 4, we conducted the same experiment but with professional software developers in industry and compared the result to the true implementation time, as implemented later by their colleagues. Experiment 5 was conducted in a field-setting using the industry teams' own requirements. In Experiment 6, we used the same teams' own backlogs and sprints with requirements that they themselves implemented afterwards. We also collected qualitative data through interviews asking the teams why they thought the estimations were inaccurate. 

Table \ref{Tab:SumofExp} provides a summary of the six experiments (Experiments 1--6), including subjects, number of requirements and obsolete requirements, type of replication according to the taxonomy by \citet{baldassarre2014replication}, and the reason for conducting the experiment. The setup and design of each experiment is described in detail in Section \ref{sec:DesandExpMtrl} while the subjects and the selection of subjects are described in Section \ref{sec:subjects}.

\begin{table*}[!t]
\renewcommand{\arraystretch}{1.3}
\caption{Summary of the setting for each experiment.}
\begin{center}
\begin{tabular}{|p{0.12\linewidth}|p{0.12\linewidth}|p{0.15\linewidth}|p{0.20\linewidth}|p{0.20\linewidth}|}
\hline
\textbf{Experiment} & \textbf{Subjects} & \textbf{\# req / obsolete req }  & \textbf{Type of replication}  & \textbf{Reason}\\
\hline
1 & 150 Bachelor's students & 4-5 / 0-1 & Exact internal replication of Experiment 0 & To see if the results from Experiment 0 still holds\\
\hline
2 & 149 Bachelor's students & 4-5 / 0-1 & Exact internal replication of Experiment 0 and 1 & To see if the results from Experiments 0 and 1 still holds\\
\hline
3 & 60 Master's students & 8-10 / 0-2 & The same design as in Experiment 2, otherwise a differentiated replication & To study if twice as many requirements and obsolete requirements would influence the estimates\\
\hline
4 & 75 industry practitioners from two companies& 8-10 / 0-2 & The same design as in Experiment 2, otherwise a differentiated replication  & To investigate if the results from Experiments 1--3 would hold true for practitioners in industry using real requirements from their companies\\
\hline
5 & 27 industry practitioners from three companies& 10-22 / 0-5 & Similar structure and design as in Experiment 4, but a conceptual replication & To see if the effect exists when practitioners estimate requirements from their own context.\\
\hline
6 & 27 industry practitioners from three companies& 139-304 / 0-60 & Based on Experiment 5, but a conceptual replication & To see if the effect exists when teams estimate their own requirements.\\
\hline
\end{tabular}
\end{center}
\label{Tab:SumofExp}
\end{table*}

\subsection{Design and Experimental Material}\label{sec:DesandExpMtrl}


The aims of \textbf{Experiments 1 and 2} were the same as for Experiment 0, i.e. to see if obsolete requirements explicitly stated to be excluded from the effort estimation have an impact of the size of the estimates. The reason for performing Experiments 1 and 2 was to investigate if the results from Experiment 0 still holds by exactly replicating Experiment 0 as reported in \citet{gren2017possible}. Thus, the design of Experiments 1 and 2 was exactly the same as for Experiment 0 (i.e.\ an internal replication). The first and second experiments had a sample size of 150 and 149 students respectively. In Experiments 1 and 2, three different tasks (A, B, and C) were designed and randomly distributed to three groups of students (Group A - performing Task A, Group B performing Task B, and Group C performing Task C) in the same class. The groups were not overlapping, i.e., in Experiment 1, 50 students performed Task A, 50 students performed Task B, and 50 performed Task C. The first group (Group A) was to estimate how long time, in weeks, it took to implement four requirements. The four requirements were:

\begin{itemize}
  \item \textit{R1: The system shall receive uncompressed data and shall compress and save the data to desired JPEG size}
  \item \textit{R2: The maximum delay from a call answer is pressed to opened audio paths is X ms}
  \item \textit{R3: The system shall have support for Time Shift (playback with delay)}
  \item \textit{R4: The system shall have a login function that consists of a username and a password}
\end{itemize}
 
Group B was given the same four requirements as Group A plus one extra added requirement, hence Group B had five requirements to estimate the total effort it would take to implement the requirements. The fifth requirement was: 

\begin{itemize}
\item \textit{R5: It shall be possible to dedicate a host buffer in RAM that is configurable between X to Y MB for HDD}
\end{itemize}

Since all of the five requirements were from one of our industrial partners, we had to replace the real values with “X and Y” in this paper due to confidentiality reasons. However, the students had the real values in their tasks. Group C was given the same five requirements as Group B, but was instructed to leave the last requirement (R5) out of the estimation.
 
Both Experiment 1 and 2 were conducted during one lecture in a mandatory course.  The students were given an introduction followed by a problem description. Then, a pre-questionnaire was handed out to the students to collect the students experiences and knowledge in relation to the English language, experience from software development in industry, and experience in effort estimation. After the pre-questionnaire was filled in by the students, the assignments and its instructions were given to the students. At this point, the students had time to read the instructions and to complete the estimation task. The effort estimation task was designed and conducted individually by the students. In total, the experiment lasted for about one hour, including introduction, explanations, pre-questionnaire, and completing the tasks. The actual time spent on the tasks, including reading the instructions and performing the estimation, was between 10-20 minutes. Since we also conducted Experiment 0, we present an analysis of all these three experiments (0, 1, and 2) jointly (see Section \ref{method1}).

The results in \citet{jorgensen} indicated that the length of the requirements specification had an impact on effort estimations, therefore it was of interest to study the degree to which twice as many requirements and obsolete requirements would influence the estimates. Thus, in \textbf{Experiment 3}, we decided to double the number of requirements for all three tasks (A, B, and C) and to conduct the experiment with a different set of students from a different university. The third experiment had a sample size of 60 students. Since the task in each group was different from the previous experiments we could not compare the result with the results from Experiments 0--2. The design of Experiment 3, including the random distribution of students into groups, was exactly the same as for Experiments 1 and 2 (i.e. a differentiated replication), except for the number of requirements and obsolete requirements, which had been doubled in size. That is, instead of using four requirements in Group A, we had eight requirements, while the number of requirements for Group B increased from five to 10 requirements. Finally, for Group C the number of requirements increased from five to 10 where the students were told to not to take the last two requirements (instead of only one as in the previous experiments) into account when performing the estimation.

Experiments 1--3 were conducted with student subjects that did not have any knowledge/expertise about the requirements, the domain or the product of which the requirements belong to, nor did they have any extensive industrial experience of software development and effort estimation. Therefore, the aim of \textbf{Experiment 4} was to investigate if the results from Experiments 1--3 would hold true for practitioners in industry using real requirements from their companies that were to be implemented in their coming sprints shortly after Experiment 4 (note that the selected requirements in Experiment 4 were not yet implemented at the time of the experiment). Experiment 4 had exactly the same number of requirements as in Experiment 3, but since the context was very different, we did not compare students' result to the result of the industry participant.  Moreover, another aspect that Experiments 1--3  do not address is the investigation of the accuracy of the estimates since we did not compare to an actual implementation effort. Therefore, when the requirements used in Experiment 4 had been implemented, we collected  the actual effort it took to implement the requirements. The fourth experiment had a sample size of 75 industry practitioners from two different companies. Experiment 4 was a differentiated replication of Experiment 3 and the design of Experiment 4 was exactly the same as for Experiment 3, except for the used requirements and having industry practitioners instead of student subjects. The main criteria used when selecting the 10 requirements were that they should be implemented in a real project after the experiment (to know the actual effort), and that the requirements should be understandable for all participating industry practitioners. Due to confidentiality reasons, the used requirements are not allowed to be revealed. Moreover, the questions asked in the pre-questionnaire differed from the ones used in Experiments 1--3. In Experiment 4, we asked questions about the subjects total years of experience in software development, total years of experience at their current company, and total years of experience with requirements engineering and effort estimation. These numbers were known for the sample as a whole and averaged out the effect of experience by randomizing the industry participants into the different group A, B, and C anyways. 

Although the subjects in Experiment 4 comprised of industry practitioners, the subjects did not estimate requirements that they were to implement. Instead, the requirements in Experiment 4 were implemented by other practitioners in the companies. Hence, the effect might only exist in contexts where an outsider, i.e. someone that will not actually implement the requirements, conduct the estimation. Therefore, \textbf{Experiment 5} looked into this aspect by being fully in the context of developers that both estimated and later implemented the requirements. Experiment 5 was conducted in a field-setting using the industry teams' own requirements. The effort estimation in Experiment 5 was based on the industry practitioners’ real requirements from their real product and sprint backlogs. The fifth experiment had a sample size of 27 industry practitioners from five complete teams at three different companies. For Experiment 5, we searched among our industrial collaboration network for software developing companies that would be interested in participating in the experiment. Three companies (hereafter named as Company C, Company D, and Company E) and five complete teams (three from Company C and one each from companies D and E, as shown in Table \ref{Tab:CompCharExp5}) were interested in the effort estimation work and decided to participate in Experiment 5. To setup and plan the experiment and to identify industry practitioners for participating in Experiment 5, we contacted three ``gate-keepers'' (one from each company).

\begin{landscape}
\begin{figure}
\includegraphics[scale=0.67]{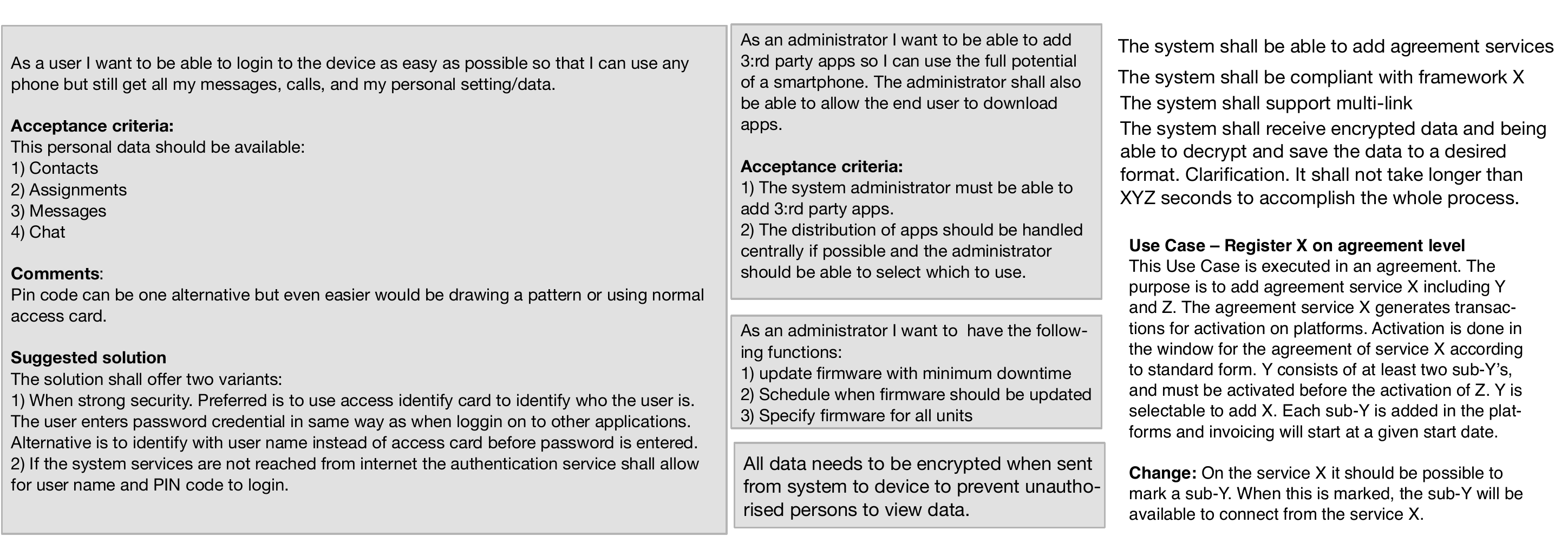}
\caption{Example of what level of details the requirements had in Experiments 5 and 6.}
\label{fig:ExReq}    
\end{figure}
\end{landscape}

Experiment 5 followed a similar structure and design as Experiment 4 (i.e. a conceptual replication), but with real requirements from the teams' real projects where the number of requirements and obsolete requirements varied. Fig. \ref{fig:ExReq} illustrates what level of details the requirements had (written as user stories, natural language requirements, and use cases) in Experiments 5 and 6. Note that the requirements in Fig. \ref{fig:ExReq} are not the real requirements that were used in Experiments 5 and 6 (due to confidentiality reasons, the used requirements are not allowed to be revealed).

For each team, the estimation effort was performed individually over two or three sprints where the number of requirements and obsolete requirements differed, both between the teams and in the sprints for each team. The reason for this difference was based on input from the ``gate-keepers'' at each company. After each sprint, the real requirements were implemented and then we collected the actual implementation effort in order to compare with the individual estimates. The main criteria used when selecting the requirements were that they should be real requirements from the team's product and\slash or sprint backlog, and that the requirements should be implemented in the coming sprint. Before the estimation of the requirements in the first sprint, the industry practitioners were given the same introduction, problem description, and pre-questionnaire as the subjects in Experiment 4. After the pre-questionnaire was filled in by the industry practitioners, the assignment (the selected real requirements from their coming sprint, which was selected by the ``gate-keepers'') and its instructions were given to the industry practitioners. This was done before the estimations of sprint 1. The industry practitioners completed the estimation work and implemented the requirements. At the beginning of sprint 2 and sprint 3, the selected requirements for each sprint were given to the industry practitioners. Again, the ``gate-keepers'' selected which requirements to include for sprint 2 and sprint 3. Please note that there was no introduction and pre-questionnaire for the second and third sprints. In total, the estimation work for each sprint lasted for about 30 minutes, while the introduction before sprint 1 lasted about 20 minutes. The ``gate-keepers'' collected the actual implementation effort from each sprint and informed the second author about the actual effort.

After conducting Experiment 5, it was not possible to decide if the effect was due to the tasks being interpreted as unrealistic. Moreover, many industry practitioners perform their effort estimations by discussion in teams, thus it was unknown if group discussions may mitigate the error. In addition, there were no details/results of how the subjects reasoned when performing the estimations where obsolete requirements were visible. All of these issues were addressed in \textbf{Experiment 6}. The sixth experiment had exactly the same subjects and companies as in Experiment 5.

\begin{landscape}
\begin{figure}
\includegraphics[scale=0.8]{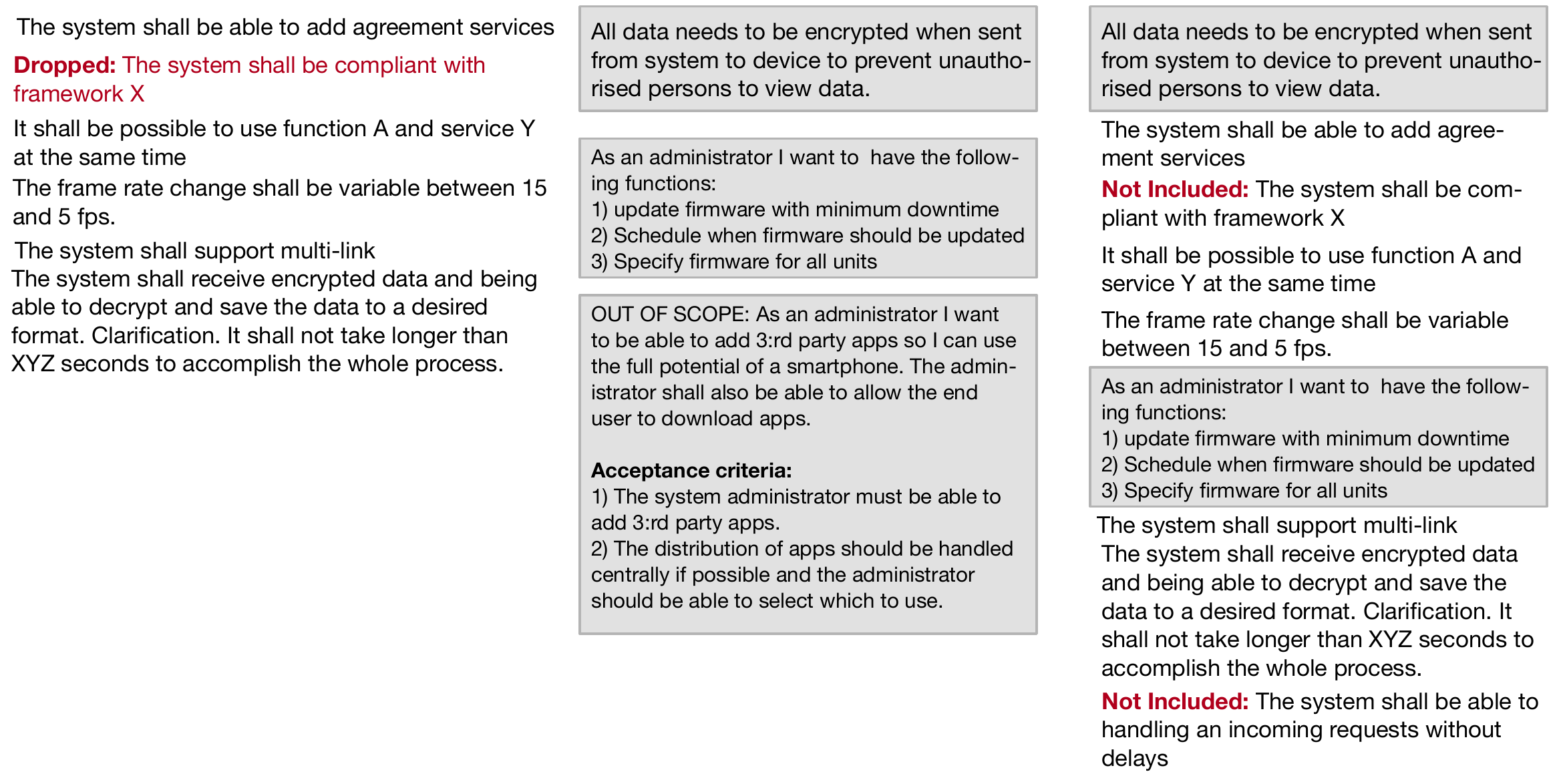}
\caption{Three examples of how obsolete requirements were marked and mixed with non-obsolete requirements in Experiments 5 and 6.}
\label{fig:ExReqandObsReq}    
\end{figure}
\end{landscape}

The purpose of Experiment 6 was to create a setup that was exactly the same as when the teams work in their daily work. Moreover, the number of requirements in the previous experiments, did not reflect the number of requirements in real projects and real sprints. Therefore, we discussed with the ``gate-keepers'' at each company about modifying (i.e. ``marking'' requirements as obsolete requirements) some real requirements in the real product backlogs for the teams, without the teams knowledge that they were still part of the study. We obtained approvals from the companies and the ``gate-keepers'' to do this in order to study the affect of obsolete requirements in real situations without the possible bias from the subjects that they are aware of being part of a study. In Experiment 6, no modifications were made to the companies or the teams processes, ways of working, how requirements end up in different backlogs, decisions-making, implementation of requirements, or how estimations were done. The only modification of the companies and the teams processes and requirements was that the ``gate-keeper'' at each company modified some of the already existing requirements in the teams’ product backlogs by ``marking'' a number/selection of requirements as obsolete as they are usually marked in their real product backlogs. For example, by stating that a requirement is ``obsolete,'' ``not included,'' ``out of scope'' or simply by marking a requirement with red color. Fig. \ref{fig:ExReqandObsReq} illustrates three examples how obsolete requirements were marked, and how they were presented to the teams together with non-obsolete requirements. Note that the requirements in Fig. \ref{fig:ExReqandObsReq} are not the real requirements that were used in Experiment 6. Each team worked in their normal product and sprint backlogs in their real projects and performing the estimations and prioritization as they normally do. That is, they looked into their product backlogs (that both contained requirements and obsolete requirements) to estimate and select which requirements should be included in the next sprint and added the selected requirements to their sprint backlog. Then, the teams implemented the requirements from the sprint backlog. All the teams had access to their product backlog, which means that they saw (and could access) all the requirements, including the added/modified obsolete requirements. What the team decided to implement in a sprint was a subset of the product backlog and discussed in the sprint planning meeting. After each sprint was completed, the ``gate-keepers'' sometimes added and/or changed the number of obsolete requirements, which always happens according to them. The number of obsolete requirements for each team and in each sprint was decided by each ``gate-keeper'' to make it as realistic as possible. That is, the researchers did not influence the percentage of obsolete requirements in the product backlogs. The used requirements in Experiment 6 had the same level of details as in Experiment 5 (see Fig. \ref{fig:ExReq}). Due to confidentiality reasons, the used requirements are not allowed to be revealed. In addition, after the requirements from the sprint backlog was implemented, the ``gate-keepers'' collected the actual implementation effort for the requirements in order to compare the actual effort with the estimations. Then the process was repeated for each sprint. In total, this process lasted for three sprints for each team. After the three sprints, the second author went back to the companies to interview the team members about their experiences. The interviews used a semi-structured approach and lasted between 10 and 30 minutes. In each interview, which was conducted face-to-face at each company, one industry practitioner and the second author participated. During the interviews, notes were taken. Experiment 6 lasted for three sprints for each team, thus the total time (in weeks) for Experiment 6 was between six and nine weeks (depending on the sprint length for each team).

\subsection{Subjects}\label{sec:subjects}
\textbf{Experiment 1} comprised of Bachelor's students from the course Software Engineering Process --- Economy and Quality at Lund University, Sweden. The course was a mandatory course for third year students offered to students at the Computer Science and Information program. In total, 150 students participated in Experiment 1, which was conducted after Experiment 0. As in Experiment 0, we distributed a pre-questionnaire. The results from the pre-questionnaire in Experiment 1 showed a small variation in the English language, ranging from ``very good knowledge'' to ``fluent.'' Out of the 150 subjects, six had industrial experience of software development (between four and eight months), and five of these six subjects had about one month experience of effort estimation.

The subjects in \textbf{Experiment 2} were Bachelor's students from the course Software Engineering Process --- Soft Issues at Lund University, Sweden. The course was a mandatory course for second year students offered to students at the Computer Science and Information program. In total, 149 students participated in Experiment 2. Experiment 2 was conducted in the same year as Experiment 1. The pre-questionnaire (the same as in Experiment 1) showed that the students' English language knowledge varied between ``good knowledge'' to ``fluent.'' Only one student had experience from software development in industry (about five months experience), while none of the students in Experiment 2 had any experience of effort estimation.


The subjects in \textbf{Experiment 3} were Master's students from the course Requirements Engineering at Chalmers $|$ University of Gothenburg, Sweden. The course was a mandatory Master's-level course for students at the educational Master's programs of Software Engineering and Interaction Design and Technologies. In total, 60 students participated in Experiment 3. Experiment 3 was conducted after after Experiment 2. In Experiment 3, the result of the pre-questionnaire revealed a variation in the English language knowledge, ranging from ``good knowledge'' to ``fluent.'' For experiences from software development in industry, most of the students reported no experience at all (52 out of 60), and for experience of effort estimation 53 out of 60 students reported no experience. For the students that reported that they had experiences from software development in industry, the experiences varied between five months up to one year. The reported experiences of effort estimation was about one month.

The subjects in \textbf{Experiments 4, 5, and 6} were industry practitioners from five different companies. For the industrial subjects, we contacted one ``gate-keeper'' at each of the five companies. The ``gate-keepers'' identified industry practitioners that (s)he thought were the most suitable and representative of the company to participate in this study, i.e.\ the ``gate-keepers'' knew that the research was about effort estimation of requirements and were to select participants that perform such work within the organization. That is, the researchers did not influence the selection of the industry practitioners, nor did the researchers have any personal relationship to the industry practitioners. The ``gate-keepers'' selected software professionals that work with requirements engineering and perform estimation work. None of the industry practitioners were students working part-time at the companies. All of the industry practitioners were fully employed by their respective company at the time of the experiments. For Experiment 4, the ``gate-keepers'' identified individual industry practitioners, while for Experiments 5 and 6, instead of identifying individual industry practitioners the ``gate-keepers'' identified complete teams that work together at the companies in their real projects. Moreover, in Experiments 5 and 6, the ``gate-keepers'' selected industry practitioners that, in addition to working with requirements engineering and perform estimation work, also were responsible for implementing the requirements. In the industrial settings (Experiments 4-6), the pre-questionnaire asked questions about the subjects total years of industrial experiences in software development, total years at their current company, and total years of experiences of requirements engineering and effort estimation. 

In total, 75 industry practitioners participated in \textbf{Experiment 4}, 21 from Company A and 54 from Company B. For the industry practitioners from Company A, the subjects had between 2 to 15 years of professional experience in software development, between 1 and 15 years of experiences at Company A, between 2 and 9 years of experiences in requirements engineering, and 2 to 6 years of experiences with effort estimations. For the industry practitioners from Company B, they had between 1 and 25 years of professional experience in software development, between 1 and 17 years of experiences at Company B, between 1 and 21 years of experiences in requirements engineering, and 1 to 18 years of experiences with effort estimations. The two companies, both from the telecommunication domain, varied in size around 250 employees at Company A and more than 2,700 employees at Company B. Both companies used agile development methods where Company A performed effort estimations individually while Company B performed effort estimations in teams. Both companies used hours as their effort estimation unit. More details about the two companies are not revealed for confidentiality reasons.

In total, 27 industry practitioners from five teams at three different companies participated in \textbf{Experiment 5 and 6}, as shown in Table \ref{Tab:SubjCharExp5}. From Company C, 18 industry practitioners from three teams participated in Experiment 5. The industry practitioners from Company C had between 3 and 15 years of professional experience at Company C and between 3 and 20 years of professional experience in software development. From Company D, four industry practitioners from one team participated in the experiment. The industry practitioners from Company D had between 4 and 10 years of professional experience in software development and between 3 and 6 years of experiences at Company D. From Company E, 5 industry practitioners from one team participated in Experiment 5. The industry practitioners from Company E had between 1 and 8 years of professional experience at Company E and between 1 and 15 years of professional experience in software development.


\begin{table*}[!t]
\renewcommand{\arraystretch}{1.3}
\caption{Industry subject characteristics - Experiments 5 and 6.}
\begin{center}
\begin{tabular}{|p{0.10\linewidth}|p{0.05\linewidth}|p{0.20\linewidth}|p{0.20\linewidth}|p{0.20\linewidth}|}
\hline
\textbf{Company} & \textbf{Team} & \textbf{Subject/Role}  & \textbf{Number of years of experience in current company} & \textbf{Number of years of experience in software development}\\
\hline
C & C.1 & Developer 1 & 6 & 10\\
 & & Developer 2 & 8 & 12\\
 & & Developer 3 & 6 & 10\\
 & & Developer 4 & 4 & 4\\
 & & Product owner & 5 & 15\\
 & & Senior engineer & 8 & 8\\
 & C.2 & Developer 1 & 5 & 7\\
 & & Developer 2 & 3 & 3\\
 & & Developer 3 & 3 & 3\\
 & & Product owner & 15 & 19\\
 & & Software designer & 11 & 20\\
 & C.3 & Developer 1 & 8 & 13\\
 & & Developer 2 & 9 & 10\\
 & & Developer 3 & 5 & 5\\
 & & Product owner & 9 & 9\\
 & & Senior engineer & 4 & 10\\
 & & Software designer & 6 & 9\\
 & & Software architect & 7 & 16\\
 \hline
D & D.1 & Developer 1 & 4 & 4\\
 & & Developer 2 & 3 & 5\\
 & & Developer 3 & 4 & 5\\
 & & Project manager & 6 & 10\\
 \hline
 E & E.1 & Developer 1 & 2 & 2\\
 & & Developer 2 & 1 & 1\\
 & & Developer 3 & 2 & 5\\
 & & Project manager & 8 & 15\\
 & & Product owner & 1 & 5\\
\hline
\end{tabular}
\end{center}
\label{Tab:SubjCharExp5}
\end{table*}

The three companies (Company C, D, and E) are in different domains and varied in size in terms of number of requirements in their backlogs, product backlogs, and sprint backlogs (as shown in Table \ref{Tab:CompCharExp5}). Company C, from the Telecommunication domain, had about 10,000 requirements in their backlog. For the three teams (C.1, C.2 and C.3 in Table \ref{Tab:CompCharExp5}) from Company C, the product backlogs varied between 150 and 400 requirements, while the sprint backlogs varied between 5 and 30 requirements. For all three teams, the sprint length was two weeks. In Team C.1, the requirements are specified using natural language (about 75\% of the requirements) and user stories (about 25\%). In Team C.2, all of the requirements are specified as natural language requirements. Team C.3 used four different specification techniques for their requirements, about 40\% of the requirements were specified using natural language and 40\% as use cases. About 15\% of the requirements were specified as user stories and 5\% as sequence diagrams. For Company D, which is a consultancy company, the product backlog had about 10,000 requirements. The product backlog for Team D.1 from Company D had between 100 and 400 requirements, while their sprint backlog varied between 15 and 20 requirements. The sprint length for Team D.1 was two weeks. Team D.1 specified all of their requirements as natural language requirements. For Company E, from the consumer electronics domain, their backlog had about 4,000 requirements, while the product backlog for Team E.1 varied between 140-180 requirements. Team E.1's sprint backlog varied between 10 and 20 requirements and the length of their sprint was three weeks. Team E.1 specified all of their requirements as user stories.

All three companies (Company C, D, and E) used agile development methods where the effort estimations were performed in teams using hours as the estimation unit at all five teams. More details about the three companies and the five teams are not revealed due to confidentiality reasons. 

\begin{table*}[!t]
\renewcommand{\arraystretch}{1.3}
\caption{Company characteristics - Experiments 5 and 6}
\begin{center}
\begin{tabular}{|p{0.10\linewidth}|p{0.10\linewidth}|p{0.10\linewidth}|p{0.05\linewidth}|p{0.15\linewidth}|p{0.15\linewidth}|p{0.10\linewidth}|}
\hline
\textbf{Company} & \textbf{Domain} & \textbf{\# requirements in backlog}  & \textbf{Team} & \textbf{\# requirements in product backlog} & \textbf{\# requirements in sprint backlog} & \textbf{Sprint length (in weeks)}\\
\hline
C & Telecom & 10,000 & C.1 & 200--300 & 15--20 & 2\\
&  &  & C.2 & 150--200 & 10--30 & 2\\
&  &  & C.3 & 200--400 & 5--15 & 2\\
\hline
D & Consultant & 10,000 & D.1 & 100--400 & 15--20 & 2\\
\hline
E & Consumer electronics & 4,000 & E.1 & 140--180 & 10--20 & 3\\
\hline
\end{tabular}
\end{center}
\label{Tab:CompCharExp5}
\end{table*}

\section{Results}\label{sec:results}
In this section we first present the results from the separate analyses conducted and then we analyze all of them together. 

\subsection{Experiments 0, 1, and 2}\label{method1}
\begin{figure}
\includegraphics[scale=0.4]{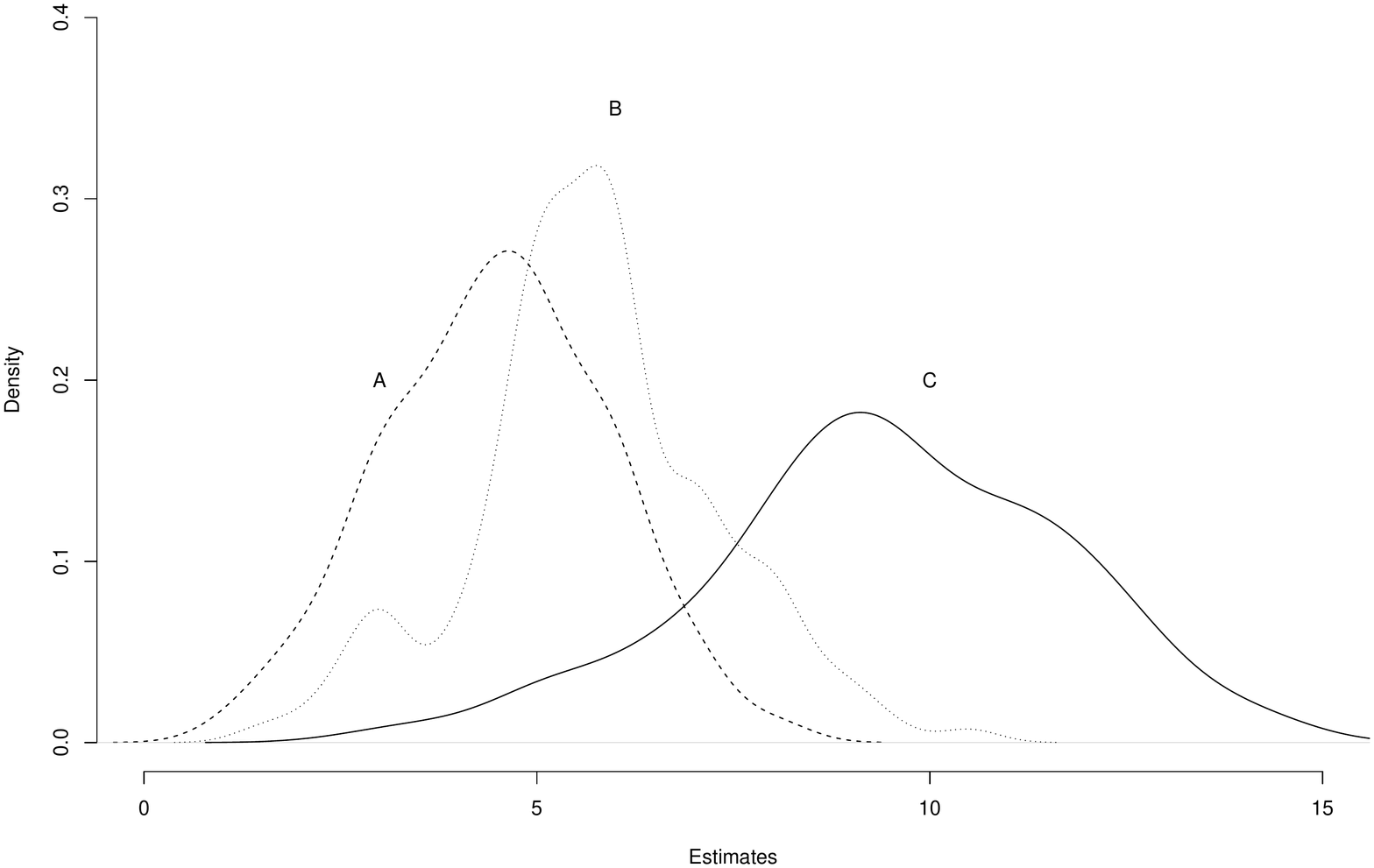}
\caption{Density plots of the raw data of the estimates for the different groups in Experiments 0 to 2.}
\label{fig:raw}       
\end{figure}

We start by plotting our raw data of the estimations obtained for each of the Groups A, B, and C. In Fig.~\ref{fig:raw}, we can see that we have quite normally distributed raw data and there seems to be a difference in that $A<B<C$. 
The likelihood functions and our weakly informative priors \citep{bernardo1975non} used when the first data was analyzed were the following:
{\footnotesize
\begin{IEEEeqnarray}{rCl}
\mathrm{Estimate}_i & \sim & \mathrm{Normal}(\mu_i, \sigma)\\
\mathrm \mu_i & = &\beta_A \mathrm{A}_i + \beta_B \mathrm{B}_i + \beta_C \mathrm{C}_i\\
\beta_A & \sim & \mathrm{Normal}(0,1) \\
\beta_B & \sim & \mathrm{Normal}(0,1)\\
\beta_C & \sim & \mathrm{Normal}(0,1)\\
\sigma & \sim & \mathrm{HalfCauchy}(0,5)
\end{IEEEeqnarray}}
Note that we have a model without any intercept (3). We could use an intercept as Group A, but if we model like this, we get much more straightforward output from brms (see Supplementary Material). The priors above need some explanation. The response variable is always assumed to be Gaussian (i.e.\ normally distributed) in linear regression \citep{McElreath2016sra} which is why our estimate variable is assumed to be Gaussian with a $\mu_i$, and $\sigma$ (2). Fig.~\ref{fig:raw} also supports this claim. We obtain a posterior distribution for each of the groups, which makes them very easy to compare (4--6). When using BDA and explicitly defining our statistical model like this, it makes it possible to directly observe our hypothesis about the experiment since we could use our subjective knowledge as priors in the statistical model. In our case, not much was known about the prior distribution; however, our assumption was that the estimates given for group A should be larger than zero and not have more extreme values than 100 (the max value obtained in our data was 14.5), which will cover extreme values. It is hard to assess how a model behaves without simulating output, which is done in a sensitivity analysis (see Supplementary Material). In brief, we tested different models and chose the one above since the simulated values of the estimate were much larger than 100. We used a standard weak informative prior for sigma (7), the Half-Cauchy prior with a standard deviation of 5 \citep{gelman2008weakly}.

\begin{figure}
\includegraphics[scale=0.4]{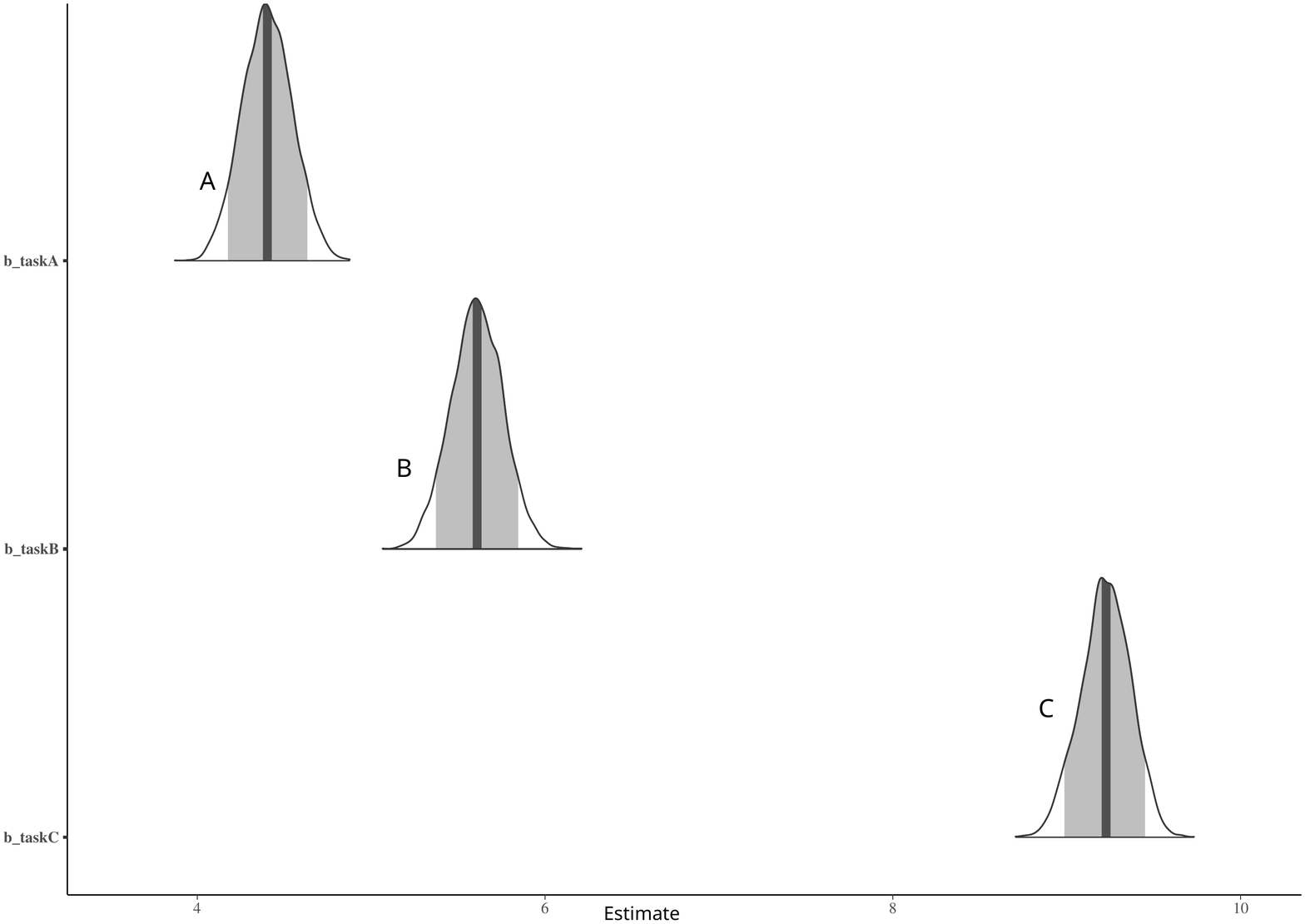}
\caption{Sampled posterior distributions in Experiment 0--2 for groups with median and 95\% credible interval (note that the sigma is not included).}
\label{fig:posteriors1}       
\end{figure}

Fig.~\ref{fig:posteriors1} shows the sampled posterior distributions, which confirms the result that was previously published, i.e.\ there is a significant difference between all the three mutually exclusive estimation groups (A: 4 requirements, B: The same 4 requirements but a fifth one added, C: The same 5 requirements as in B but the fifth was marked ``Please note that requirement 5 should not be implemented''). 
Table~\ref{tab:1} shows the parameter statistics for each Group A, B, and C. We see that all the Groups are different and we obtained much higher estimates in Group C where one requirement was marked as obsolete.

\begin{table}
\caption{Means and 95\% credible interval for for the groups parameters and the sigma used in the likelihood model.}
\label{tab:1}       
\begin{tabular}{llll}
\hline\noalign{\smallskip}
 & Mean  & l-95\% CI & u-95\% CI\\
\noalign{\smallskip}\hline\noalign{\smallskip}
Group A & 4.43  & 4.15 & 4.70\\
Group B & 5.87  & 5.59 & 6.15\\
Group C & 9.41  & 9.13 & 9.69\\
Sigma & 1.83  & 1.72 & 1.95\\
\noalign{\smallskip}\hline
\end{tabular}
\end{table}

By simply looking at Figure~\ref{fig:posteriors1} or reading Table~\ref{tab:1}, we see that all the groups were significantly different from each other too since almost no values even overlap. However, a measurement of effect size overall was important to calculate. The Bayesian $R^2$ was 53.8\%, which mean that around 54\% of the variance in estimations can be explained by Group, which is very high considering so many other confounding factors when people make estimations of requirements. By this we mean all the unexplained variance present in a behavioral context that should be averaged out instead of blocked. This is why effect sizes in psychological science are considered high with quite low percentage of explained variance \citep{cohen}, because they are not low in a complex system.

\subsection{Experiment 3}\label{exp3}
Since it was not possible to know how the longer requirements specifications with 8 and 10 requirements would affect the estimations, we used weak priors again for the third experiment, i.e.\ we started our data analysis with exactly the same model and priors as in the previous data analysis. Based on the results from the previous experiments, we could have assumed group C to be larger than A and B; however, with the new and longer requirements specification we opted to be very conservative and careful regarding the effect of C.

\begin{figure}
\includegraphics[scale=0.4]{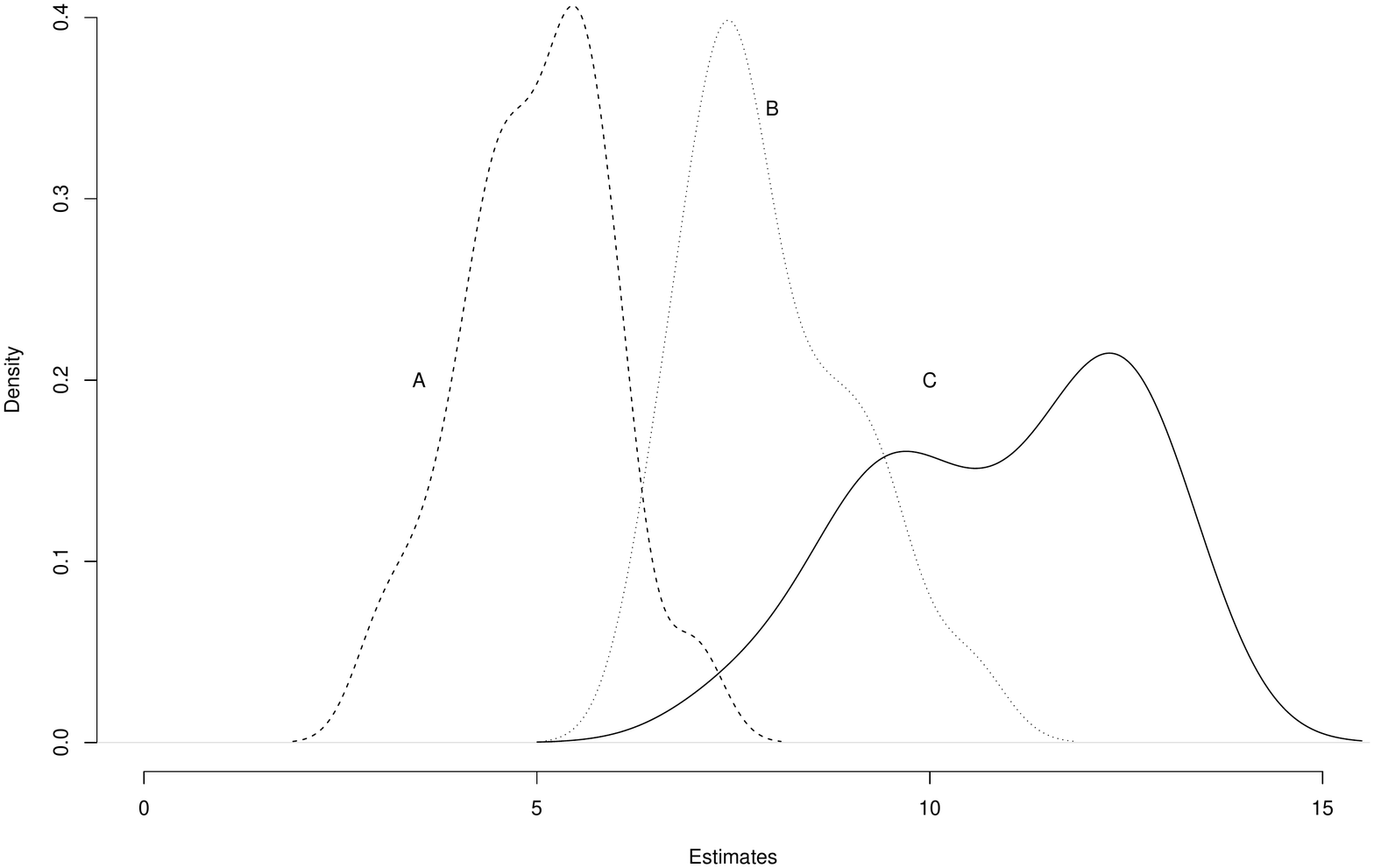}
\caption{Raw data for the different groups in Experiment 3.}
\label{fig:raw2}       
\end{figure}

We start again by plotting our raw data of the estimations obtained for each of the Groups A, B, and C. In Fig.~\ref{fig:raw2}, we can see that we have quite normally distributed raw data and there seems to be a difference in that $A<B<C$, just like before. 

We updated our model from the previous experiments into a Log-Normal distribution due to our sensitivity analysis (see Supplementary Material). 

\begin{figure}
\includegraphics[scale=0.4]{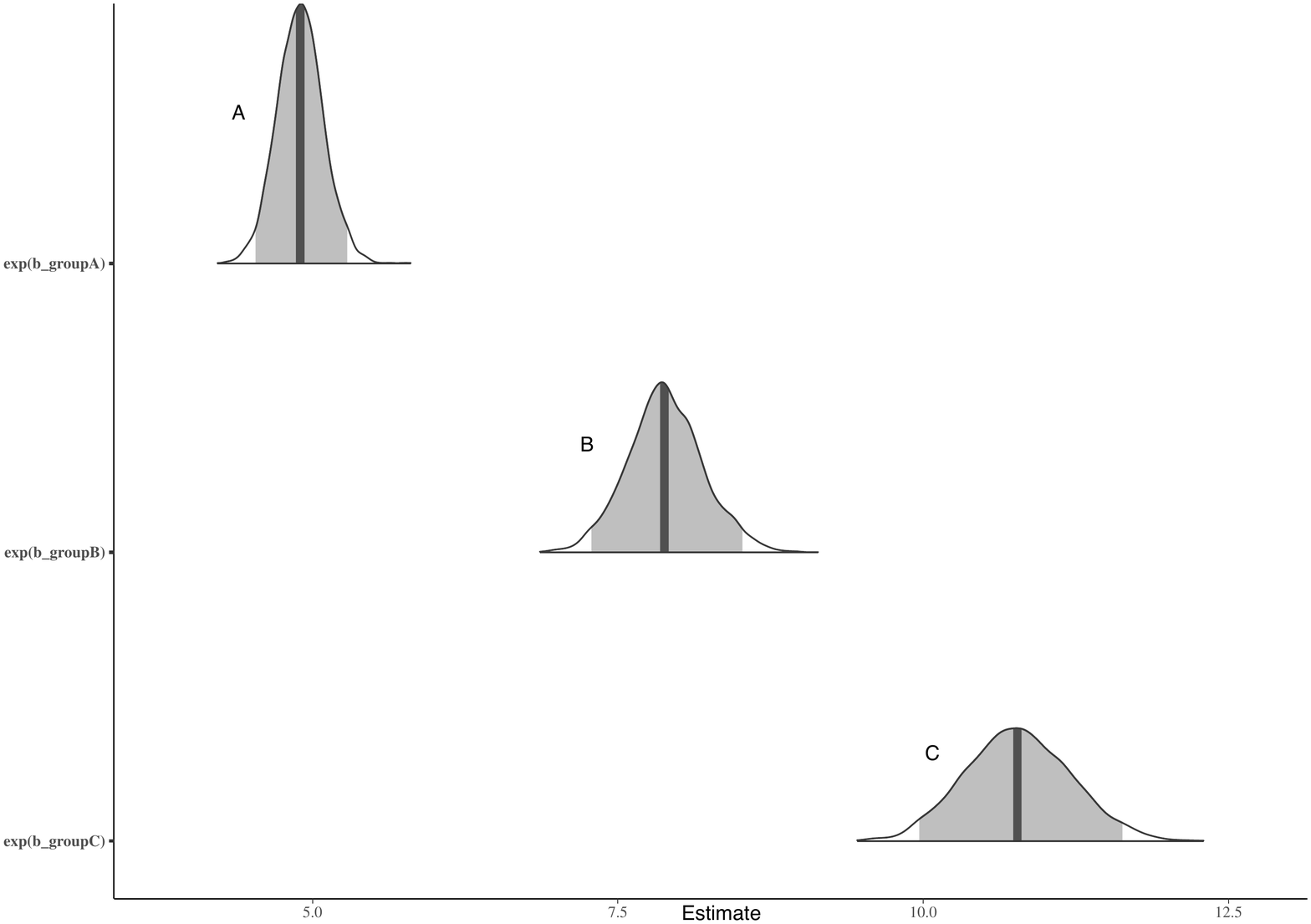}
\caption{Sampled posterior distributions in Experiment 3 for groups with median and 95\% credible interval (note that the sigma is not included).}
\label{fig:4}       
\end{figure}

{\footnotesize
\begin{IEEEeqnarray}{rCl}
\mathrm{Estimate}_i & \sim & \mathrm{LogNormal}(\mu_i, \sigma)\\
\mathrm \mu_i & = &\beta_A \mathrm{A}_i + \beta_B \mathrm{B}_i + \beta_C \mathrm{C}_i\\
\beta_A & \sim & \mathrm{Normal}(0,1) \\
\beta_B & \sim & \mathrm{Normal}(0,1)\\
\beta_C & \sim & \mathrm{Normal}(0,1)\\
\sigma & \sim & \mathrm{HalfCauchy}(0,5)
\end{IEEEeqnarray}}
Fig.~\ref{fig:4} shows the sampled posterior distributions, and Table~\ref{tab:4} shows the parameter for each groups with a connected 95\% credible interval. As we can see, the estimations for the groups, all the estimates increase and we see a similar pattern as in the previous experiments. The true implementation time for students was not known; however, it is expected that A have increased simply because more requirements should take more time to implement. Our main conclusion is still that the pattern of obtaining even larger estimates when told to exclude requirements still holds.

\begin{table}
\caption{Means and 95\% credible interval for for the groups parameters and the sigma used in the likelihood model.}
\label{tab:4}       
\begin{tabular}{llll}
\hline\noalign{\smallskip}
 & Mean & l-95\% CI & u-95\% CI\\
\noalign{\smallskip}\hline\noalign{\smallskip}
Group A & 4.90  & 4.53 & 5.26\\
Group B & 7.85  & 7.32 & 8.50\\
Group C & 10.80  & 9.97 & 11.59\\
Sigma & 1.19  & 1.15 & 1.23\\
\noalign{\smallskip}\hline
\end{tabular}
\end{table}

In the case of the third experiment our Bayesian $R^2 = 0.72$, which means that around 75\% percent of the variation in the estimations can be explained by which group (A, B, or C) the subjects were part of. We interpret this effect size as extremely high. 

\subsection{Experiment 4}\label{method4}
Experiments 0--3 were conducted with student subjects that did not have any knowledge\slash expertise about the requirements, the domain or the product of which the requirements belong to, nor did they have any extensive industrial experience of software development and effort estimation. These issues were addressed in Experiment 4.

Since this is the first experiment in industry, the analysis used the same weak prior knowledge as before. One of the biggest threats to the previous experiments was that it could be seen as a toy problem that would not exist in the real world where estimations are conducted. Hence, weakly informative priors were used again. 

\begin{figure}
\includegraphics[scale=0.4]{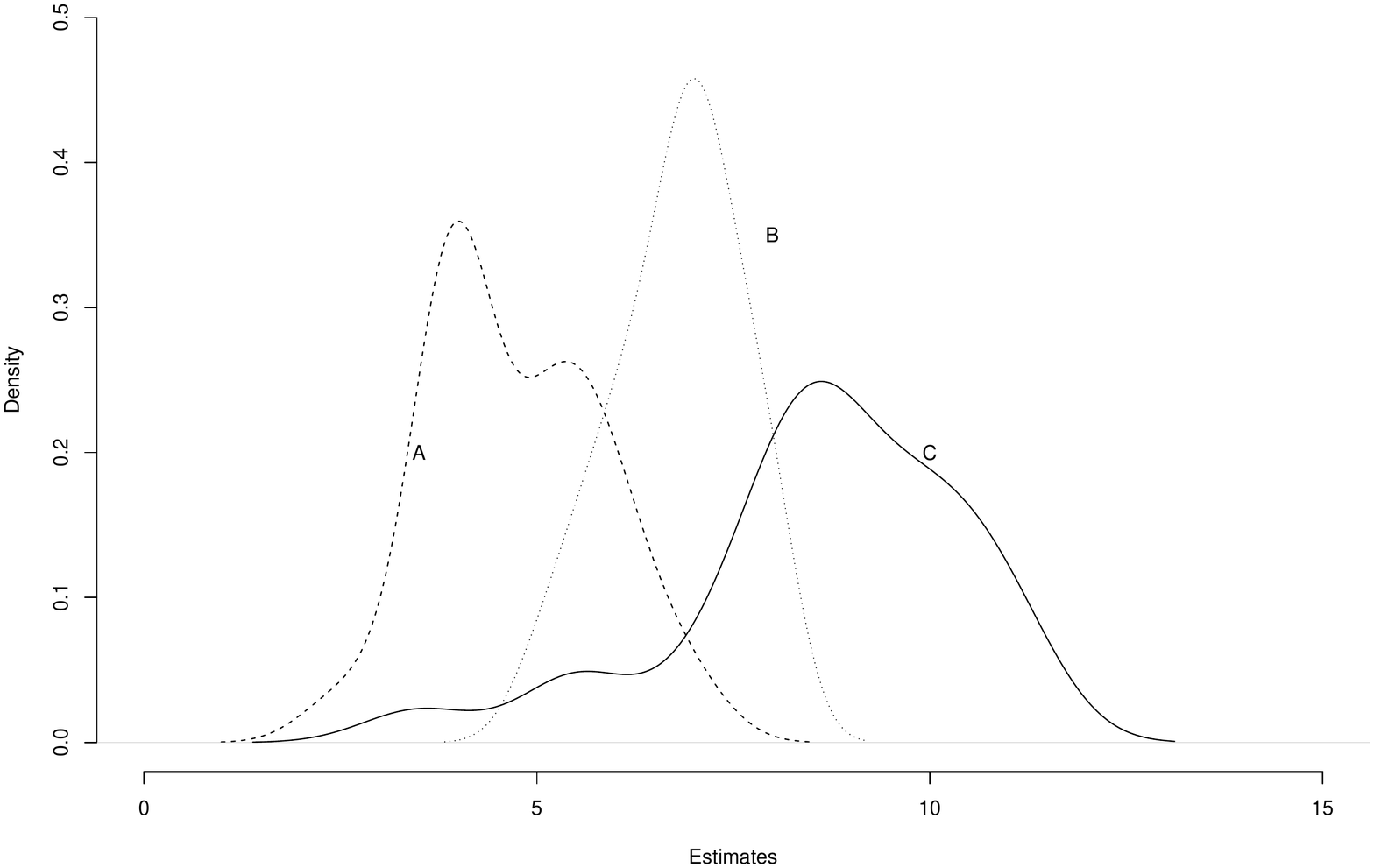}
\caption{Raw data for the different groups in Experiment 4.}
\label{fig:raw3}       
\end{figure}

We start, as always, by plotting our raw data of the estimations obtained for each of the Groups A, B, and C. In Fig.~\ref{fig:raw3}, we can again see that we have quite normally distributed raw data and there seems to be a difference in that $A<B<C$. 
For the same reason as in previous experiment, we use a Log-Normal distribution due to our sensitivity analysis (see Supplementary Material). 
{\footnotesize
\begin{IEEEeqnarray}{rCl}
\mathrm{Estimate}_i & \sim & \mathrm{LogNormal}(\mu_i, \sigma)\\
\mathrm \mu_i & = &\beta_A \mathrm{A}_i + \beta_B \mathrm{B}_i + \beta_C \mathrm{C}_i\\
\beta_A & \sim & \mathrm{Normal}(0,1) \\
\beta_B & \sim & \mathrm{Normal}(0,1)\\
\beta_C & \sim & \mathrm{Normal}(0,1)\\
\sigma & \sim & \mathrm{HalfCauchy}(0,5)
\end{IEEEeqnarray}}

\begin{figure}
\includegraphics[scale=0.4]{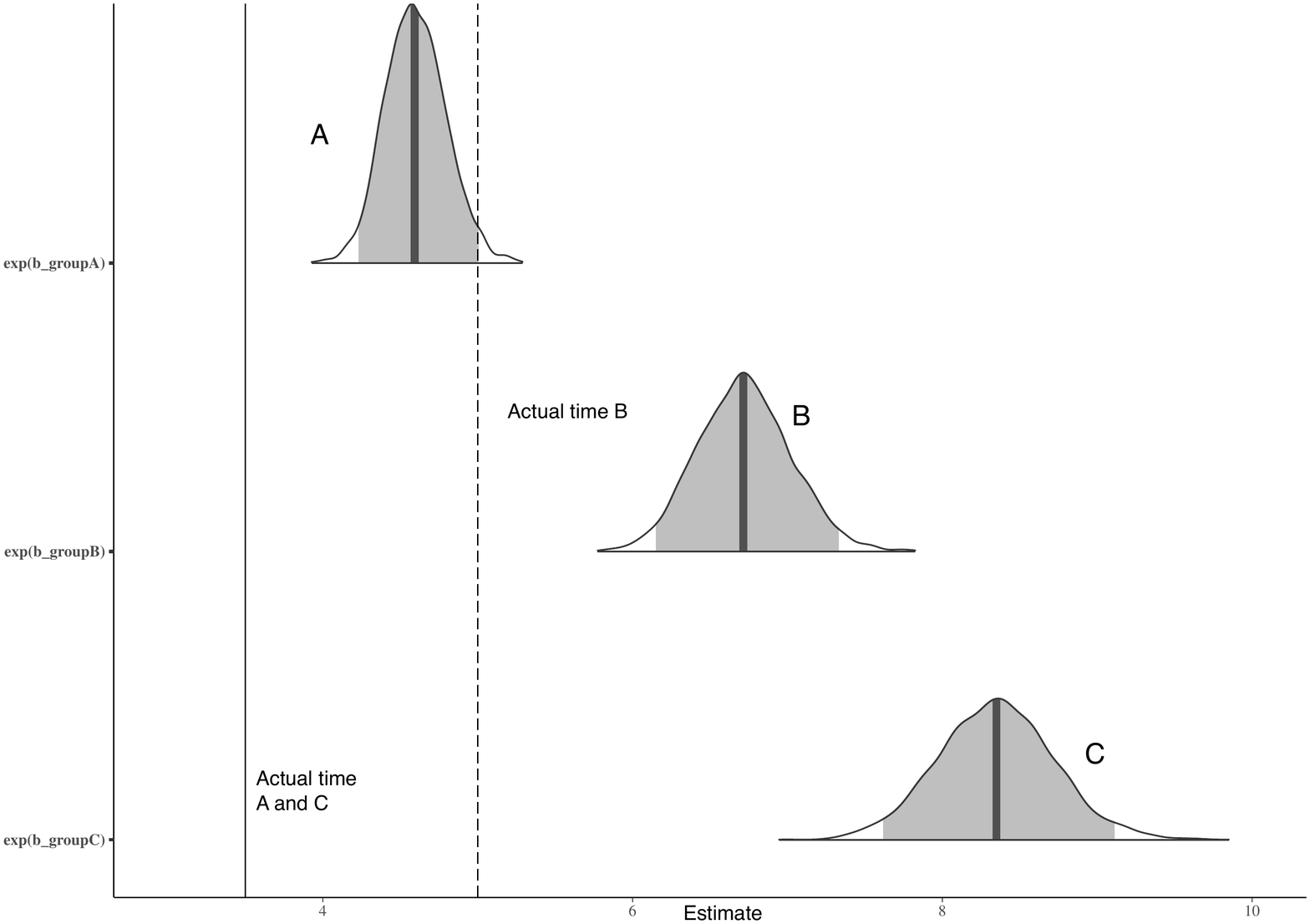}
\caption{Sampled posterior distributions in Experiment 4 for groups with median and 95\% credible interval and the two actual implementation times (the left one for A and C, and the right dashed line for B). Note that the sigma is not included.}
\label{fig:5}       
\end{figure}

The results of the experiment conducted in an industrial setting showed the same pattern again. Table~\ref{tab:5} shows the means, standard deviations, and credible interval just like in the previous experiments. Fig.~\ref{fig:5} shows the posterior distribution including two lines. The left line represents the actual implementation time for task A (3.5 weeks), and the right line (dashed) represents the actual implementation time for tasks B and C (5 weeks). We can see that in all cases the practitioners overestimated the implementation times. However, the over-estimations in A are lower (around 1.4 weeks) as compare to the estimates of more requirements in B (almost 2 weeks). The worst over-estimations were due to the marking of two requirements as obsolete increased the overestimation to close to 4 weeks.

\begin{table}
\caption{Means and 95\% credible interval for for the groups parameters and the sigma used in the likelihood model.}
\label{tab:5}       
\begin{tabular}{lllll}
\hline\noalign{\smallskip}
 & Mean & l-95\% CI & u-95\% CI\\
\noalign{\smallskip}\hline\noalign{\smallskip}
Group A & 4.62  & 4.22 & 5.00\\
Group B & 6.69  & 6.17 & 7.31\\
Group C & 8.33  & 7.61 & 9.12\\
Sigma & 1.25  & 1.21 & 1.30\\
\noalign{\smallskip}\hline
\end{tabular}
\end{table}

In the case of the fourth experiment $R^2 = 0.527$, which means that around 53\% percent of the variation in the estimations can be explained by which group (A, B, or C) the subjects were part of. We interpret this effect size as high again.

\subsection{Summary of Experiments 1--4}
We have now analyzed the first four experiments and can conclude that the fact that obsolete requirements have an effect of the estimations is clear (Experiments 1 and 2). From Experiment 3, the results show that the same effect was found using a twice as big requirements specifications including twice as many obsolete requirements. However, from the student experiments (Experiments 1--3) it is not possible to know if the students over- or under-estimated. From Experiment 4, the results show that the effect existed in industry where practitioners estimated real requirements later implemented by someone else at the company, and that it resulted in a gross over-estimation. 

The found effect sizes were 0.54, 0.75, and 0.54. Since we opted to not use any knowledge between these three sets of experiments (only between 0, 1, and 2, which led us to analyze all experimental data together), we need to be careful when comparing them or even averaging the effect. All the results show is that the effect exists and is large, even larger for larger requirements specifications and lower again in an industrial setting. The effect might only exist in contexts where an outsider, i.e.\ someone that will not actually implement the requirements, conduct the estimation. This was addressed in Experiment 5.

Based on the results until this point, it would be good to create a model that can predict the over-estimations by knowing the percentage of obsolete requirements. Unfortunately, only a small subset of our data includes any information of the true implementation time and the percentage of obsolete requirements. More specifically, only Group C in Experiment 4 include that information. The data from Experiment 4 (closest to a real setting) show that the true implementation time for Group C was 5 weeks (see Figure~\ref{fig:5}). In Experiment 6, we collected more data of that kind.

\subsection{Experiment 5}\label{method5}
Experiment 5 did not include a large enough sample from different groups who partly estimate the same requirements (only three teams from Company C). Therefore, it is not possible to assess the different levels of the effect much further. However, this was not the main aim. The aim was instead to see if obsolete requirements have a similar effect of distorting estimates when practitioners themselves estimate requirements from their own work. In Experiment 5 all participants conducted the estimations individually, and we then calculated a mean value for each team, as shown in Table~\ref{Tab:Suasd}.

The three teams from Company C had a common product backlog so we tested the same requirements (but in different order and different ones marked as obsolete) with all of them before one team then implemented them. For each team, the estimation effort was performed individually in two or three sprints where the number of requirements and obsolete requirements differed, as shown in Table \ref{Tab:Step1_Exp5}. 
\begin{table*}[!t]
\renewcommand{\arraystretch}{1.3}
\caption{Number of requirements in Experiment 5.}
\begin{center}
\begin{tabular}{|p{0.05\linewidth}|p{0.07\linewidth}|p{0.16\linewidth}|p{0.15\linewidth}|p{0.19\linewidth}|p{0.15\linewidth}|}
\hline
\textbf{Team} & \textbf{Sprint} & \textbf{\# requirements}  & \textbf{\# obsolete requirements} & \textbf{total \# requirements} & \textbf{Percent obsolete reqs}\\
\hline
C.1 & 1 & 15 & 0 & 15 & 0\%\\
& 2 & 20 & 4 & 24 & 17\%\\
& 3 & 10 & 1 & 11 & 9\%\\
\hline
C.2 & 1 & 15 & 3 & 18 & 17\%\\
& 2 & 20 & 2 & 22 & 9\%\\
& 3 & 10 & 0 & 10 & 0\%\\
\hline
C.3 & 1 & 15 & 2 & 17 & 12\%\\
& 2 & 20 & 0 & 20 & 0\%\\
& 3 & 10 & 2 & 12 & 17\%\\
\hline
D.1 & 1 & 15 & 4 & 19 & 21\%\\
& 2 & 17 & 0 & 17 & 0\%\\
\hline
E.1 & 1 & 15 & 5 & 20 & 25\%\\
& 2 & 18 & 0 & 18 & 0\%\\
\hline
\end{tabular}
\end{center}
\label{Tab:Step1_Exp5}
\end{table*}
The results from Experiment 5 are shown in Table~\ref{Tab:Suasd}. The three teams from Company C partly estimated the same requirements but different ones marked as obsolete. Overall, the results show an effect of introducing obsolete requirements, cf.\ Table~\ref{Tab:Step1_Exp5} and Table~\ref{Tab:Suasd}. Without any obsolete requirements the estimations are quite accurate, but when obsolete requirements are introduced the individuals systematically conduct over-estimations. 


\begin{table*}[htbp]
\renewcommand{\arraystretch}{1.3}
\caption{Individual and team estimations of the teams' own subset of real requirements in Experiment 5.}
\begin{center}
\begin{tabular}{|p{0.05\linewidth}|p{0.40\linewidth}|p{0.10\linewidth}|p{0.10\linewidth}|p{0.10\linewidth}|}
\hline
\textbf{Team} & \textbf{Role / Team average / actual implementation and \% actual overestimation}& \textbf{Sprint 1} & \textbf{Sprint 2}& \textbf{Sprint 3}\\
\hline
C.1  & Developer 1    & 370  & 600        &465  \\
     & Developer 2    & 345  & 605        &460 \\
     & Developer 3    & 320  & 595         &470  \\
     & Developer 4    & 330  & 620        &460  \\
     & Product owner    & 350  & 645        &450 \\
     & Senior engineer    & 356  & 612         &455  \\
     & \textbf{Team average}    & 345  & 613        &460 \\
     & \textbf{Actual implementation} & 340 & 473 & 414 \\
     & \textbf{\% actual overestimation}    & 1.5\%  & 30\%         &11\%  \\
 \hline
C.2  & Developer 1    & 455  & 520        &400  \\
     & Developer 2    & 435  & 545        &390 \\
     & Developer 3    & 490  & 500         &420  \\
     & Product owner    & 415  & 495        &395  \\
     & Software designer    & 445  & 535        &415 \\
     & \textbf{Team average}    & 448  & 519        &404 \\
     & \textbf{Actual implementation} & 340 & 473 & 414 \\
     & \textbf{\% actual overestimation}    & 32\%  & 10\%         &-3\%  \\
 \hline
 C.3  & Developer 1    & 410  & 490       &510  \\
     & Developer 2    & 405  & 450        &505 \\
     & Developer 3    & 395  & 495         &530  \\
     & Product owner    & 425  & 495        &500  \\
     & Senior engineer    & 435  & 480        &515 \\
     & Software designer    & 430  & 505         &535  \\
     & Software architect     & 395  & 500        &525 \\
     & \textbf{Team average}    & 414  & 519        &517 \\
     & \textbf{Actual implementation} & 340 & 473 & 414 \\
     & \textbf{\% actual overestimation}    & 21\%  & 3\%         &25\%  \\
 \hline
 D.1  & Developer 1    & 480  & 360       & NA  \\
     & Developer 2    & 495  & 350        & NA \\
     & Developer 3    & 535  & 330         &NA  \\
     & Project manager    & 550  & 390        &NA  \\
     & \textbf{Team average}    & 515  & 358        &NA \\
     & \textbf{Actual implementation} & 369 & 351 & NA \\
     & \textbf{\% actual overestimation}    & 40\%  & 1.9\%         &NA  \\
\hline
 E.1  & Developer 1    & 776  & 484       &NA  \\
     & Developer 2    & 796  & 528       &NA \\
     & Developer 3    & 783  & 515        &NA  \\
     & Project manager    & 785  & 498        &NA  \\
     & Product owner    & 703  & 529        &NA \\
     & \textbf{Team average}    & 769  & 511        &NA \\
     & \textbf{Actual implementation} & 575 & 503 & NA \\
     & \textbf{\% actual overestimation}    & 34\%  & 1.5\%         &NA  \\
\hline
\end{tabular}
\end{center}
NA: Not Applicable, meaning teams D.1 and E.1 only performed estimations in two sprints
\label{Tab:Suasd}
\end{table*}

We now have some more individual data on both the percentage of obsolete requirements and the resulting over-estimations. The requirements specification used in Experiment 4 all included 20\% obsolete requirements and had an actual implementation time of 3.5 weeks. As can be seen in Table~\ref{Tab:Suasd}, we obtained some more data from Experiment 5. This additional data comprised of individual estimates from several sprints in a real industrial setting.

Based on all the collected data with the percentage of obsolete requirements and the over-estimations, we created a model for predicting new values using a posterior distribution and tested that model against the values obtained in Experiment 6. We have very few data point per percentage of obsolete requirements, thus it is not expected that our model would be precise. The main reason for providing it here is to show a first model that can later be trained with more data by us or other researchers. 

All the details of this model is in Supplementary Material, but we have 70 data points to use as data. The percentage of obsolete requirements was obtain by dividing the number of obsolete requirements by the total amount of requirements included. The over-estimation was obtained by dividing the provided estimate by the actual implementation time.

\subsection{Experiment 6}\label{method6}
As previously mentioned, we tweaked Experiment 5 in Experiment 6 and injected obsolete requirements into their current sprint planning without their knowledge (with permission from the ``gate-keepers''). We injected different sets of obsolete requirements over a period of three real sprints. The setup of this part of the experiment with number of requirements and obsolete ones are shown in Table \ref{Tab:Step2_Exp5}. For example, Team C.2 started with 189 requirements, of which 38 were ``marked'' as obsolete requirements, in their product backlog (Sprint 1). From the product backlog, 12 requirements (of which 0 was obsolete requirements) were selected to be included in the sprint backlog for Sprint 1, and all 12 requirements were implemented. In the beginning of Sprint 2, the product backlog contained 177 requirements of which 18 were ``marked'' as obsolete requirements. The ``gate-keeper'' in Company C changed the number of obsolete requirements in the product backlog according to how this is always done in Company C. From the product backlog, 18 requirements, including 2 obsolete requirements, were selected by Team C.2 to be included in Sprint 2. During the implementation of the selected requirements, Team C.2 realized that two obsolete requirements were selected and thus did not implement these requirements. Hence, 16 requirements were implemented in Sprint 2. Two other teams, D.1 in Sprint 3 and E.1 in Sprint 2 did also select obsolete requirements from the product backlog to their sprint backlog. In the same way as Team C.2, both these teams realized this during their sprint and did not implement the obsolete requirements.

\begin{table*}[!t]
\renewcommand{\arraystretch}{1.3}
\caption{Number of requirements in Experiment 6.}
\begin{center}
\begin{tabular}{|p{0.05\linewidth}|p{0.05\linewidth}|p{0.10\linewidth}|p{0.15\linewidth}|p{0.10\linewidth}|p{0.20\linewidth}|p{0.10\linewidth}|}
\hline
\textbf{Team} & \textbf{Sprint} & \textbf{\# requirements in product backlog}  & \textbf{\# obsolete requirements in product backlog} & \textbf{\# requirements in sprint backlog} & \textbf{\# obsolete requirements in sprint backlog} & \textbf{\# implemented requirements in sprint}\\
\hline
C.1 & 1 & 252 & 0 & 16 & 0 & 16\\
& 2 & 239 & 24 & 15 & 0 & 15\\
& 3 & 232 & 46 & 17 & 0 & 17\\
\hline
C.2 & 1 & 189 & 38 & 12 & 0 & 12\\
& 2 & 177 & 18 & 18 & 2 & 16\\
& 3 & 163 & 0 & 25 & 0 & 25\\
\hline
C.3 & 1 & 304 & 31 & 7 & 0 & 7\\
& 2 & 299 & 0 & 8 & 0 & 8\\
& 3 & 294 & 60 & 8 & 0 & 8\\
\hline
D.1 & 1 & 287 & 0 & 12 & 0 & 12\\
& 2 & 275 & 56 & 14 & 0 & 14\\
& 3 & 263 & 56 & 18 & 1 & 17\\
\hline
E.1 & 1 & 147 & 30 & 14 & 0 & 14\\
& 2 & 139 & 30 & 20 & 2 & 18\\
& 3 & 142 & 30 & 25 & 0 & 20\\
\hline
\end{tabular}
\end{center}
\label{Tab:Step2_Exp5}
\end{table*}



The data was analyzed in the same way as in Experiment 5, but based on three real sprints per team with injected obsolete requirements in some of the sprints. We calculated a predicted impact (i.e.\ overestimation) by drawing from our posterior distribution for each percentage of obsolete requirements we wanted to predict. Since it was expected the teams would to try to adjust their estimates, qualitative data was collected by interviewing the teams after each sprint. We wanted to know how the teams were reasoning about their estimation accuracy in each sprint. What was being said in the interviews was noted immediately and summarized by the second author after each interview.

Table~\ref{Tab:Exp6} shows a simplified table as compared to Experiment 5, since the unit for the estimations etc.\ were exactly the same. Table~\ref{Tab:Exp6} only shows the real sprints for each team (3 per team), the percentage of obsolete requirements they had in their product backlog when estimating the coming sprint, the estimated effort to implement the selected requirements in each sprint (see column \# requirements in sprint backlog in Table \ref{Tab:Step2_Exp5}), the actual implementation effort of the implemented requirements (see column \# implemented requirements in sprint in Table \ref{Tab:Step2_Exp5}), the predicted overestimation based on our prediction model with a 95\% credible interval, and the actual over-estimations made by the teams.

\begin{table*}[!t]
\renewcommand{\arraystretch}{1.3}
\caption{Results and predictions in Experiment 6.}
\begin{center}
\begin{tabular}{|p{0.05\linewidth}|p{0.05\linewidth}|p{0.18\linewidth}|p{0.10\linewidth}|p{0.07\linewidth}|p{0.20\linewidth}|p{0.10\linewidth}|}
\hline
\textbf{Team} & \textbf{Sprint} & \textbf{\% obsolete reqs in product backlog}  & \textbf{Estimated effort (hours)} & \textbf{Actual effort (hours)} & \textbf{Median predicted overestimation [CI]} & \textbf{\% actual overestimation} \\
\hline
C.1 & 1 & 0\% & 460 & 454 & 0\% & 1.3\%  \\
    & 2 & 10\% & 455 & 372 & 22\% [0,91] & 22\%  \\
    & 3 & 20\% & 465 & 337 & 27\% [0,94] & 38\%  \\
\hline
C.2 & 1 & 20\% & 380 & 275 & 27\% [0,94] & 38\%  \\
    & 2 & 10\% & 385 & 300 & 22\% [0,91] & 28\%  \\
    & 3 & 0\% & 520 & 355 & 0\% & 46\%*  \\
\hline
C.3 & 1 & 10\% & 540 & 395 & 22\% [0,91] & 37\%  \\
    & 2 & 0\% & 550 & 570 & 0\% & -4\%  \\
    & 3 & 20\% & 545 & 335 & 27\% [0,94] & 63\%  \\
\hline
D.1 & 1 & 0\% & 300 & 310 & 0\% & -3\%  \\
    & 2 & 20\% & 295 & 203 &  27\% [0,94] & 45\%  \\
    & 3 & 21\% & 320 & 231 & 28\% [0,91] & 39\%  \\
\hline
E.1 & 1 & 20\% & 589 & 590 & 27\% [0,94] & 47\%  \\
    & 2 & 22\% & 630 & 464 & 29\% [0,88] & 36\%  \\
    & 3 & 21\% & 600 & 590 & 28\% [0,91] & 1.7\%**  \\

\hline
\end{tabular}
\end{center}
*The team decided to add 30\% to the third sprint based on Sprints 1 and 2.
**The team changed their estimation technique for sprint 3 into longer and more in-depth discussions and being more optimistic in their estimates. However, in Sprint 3 without any obsolete requirements, they underestimated grossly and needed an extra 150 hours (in the coming sprint) to complete what they had planned.
\label{Tab:Exp6}
\end{table*}

Overall, the results show that our model's median values are more accurate for 10\% obsolete requirements than for 20\%, as shown in Table~\ref{Tab:Exp6}. However, the 95\% credible interval says that we would expect values to lie within $<0$ (we adjusted negative values to 0) to $94$, which indicates a lot of uncertainty in our predictions. With such large uncertainty interval, we could instead create a best guess based on the latest data obtained in Experiment 6. The results show that the over-estimations triggered by the obsolete requirements were systematically twice the percentage of the included obsolete requirements, i.e.\ the percentage of obsolete requirements can be multiplied by two to get an initial idea of the potential over-estimation, at least for around 10\% obsolete ones. However, we should be careful with this model since there is much uncertainty connected to it. All that can be concluded is that obsolete requirements seem to trigger a non-linear over-estimation.

\subsubsection{Qualitative results}\label{QualRes}
In order to understand the reasons for the difference between estimates vs. actual effort, we went back to the companies to interview the team members. In general, the results from the interviews show that the teams do not know why the estimates turned out the way they did. The teams were very surprised and could not explain why they over-estimated. The results from the interviews for each team are presented below.

Team C.1 did not reflect much on their over-estimations after the second sprint. They did not think it mattered even if they were a bit surprised. Their reasoning was that if this happens just once it could be due to the properties of the features. They had the same reasoning even after sprint 3, even if it had happened twice in a row then. 

Team C.2 were very surprised by the results after the first sprint that they were done with everything so fast. They discussed a bit about it, but decided that they were unlucky and this could happen from time to time. For sprint 2, Team C.2 did include obsolete requirements in their sprint backlog. Team C.2 explained that they did not check if the requirements were marked as obsolete or not, they simply included the obsolete requirements without reading if they were obsolete. After sprints 1 and 2 they had decided to add 30\% more to implement, but with no obsolete requirements in sprint 3 they still grossly overestimated the time needed for what they selected to implement.  	
	
Team C.3 explained that they did not reflect on their over-estimates at all. Team C.3 just continued as usual without any reflections or discussions.  

Team D.1 was extremely surprised after sprint 2 and brought it up in the sprint meeting where they discussed what happened; however, they could not understand what happened or why it happened. After sprint 3, they were equally surprised again and could not understand why they grossly overestimated the implementation time twice. Especially since this has not happen for Team D.1 before.

Team E.1 had the same number of obsolete requirements in all their 3 sprints. After their first sprint, team E.1 increased the number of included features for sprint 2. However, during the second sprint, team E.1 realized that they included two features that should not be implemented, i.e. the two features were marked as obsolete. Therefore, before including features for sprint 3, team E.1 double checked the features (to make sure that they were not obsolete features), had longer and more in-depth discussions, and were more optimistic in their estimates. This led to that the time it took to decide which features to include in sprint 3 took twice as long as it usually does. As a result, team E.1 included 25 features in sprint 3, but only managed to implement 20. The five features that were not implemented took another 150 hours to implement in the coming sprint.

\section{Discussion}\label{sec:disc}
Overall, the six experiments have shown that having obsolete requirements visible when estimating software development effort, has a large effect on the size of the estimations in that they increase substantially. In the the industry samples, where we also had an actual implementation time, the obsolete requirements were shown to result in gross over-estimations. The results from \textbf{Experiments 1 and 2} show that the effect is very large when students do the estimations and that their study year (second or third year at the software engineering Bachelors' program) did not effect the estimates. The results from \textbf{Experiment 3} show that the effect was even larger with students estimating more requirements and being exposed to even more obsolete ones. The main finding of \textbf{Experiment 4} was that the effect was also present in an industrial setting with the same pattern as before. Thus, the experience of software estimation or software development does not remove the effect. In addition, the results from Experiment 4 show that the effect is in form of an over-estimation and quite a large one. From \textbf{Experiment 5}, the results show that the effect also exists when practitioners estimate their own organization's requirements individually. In \textbf{Experiment 6}, the results show that the effect also appeared when teams estimate their own work for their near-future implementation work, that the team discussion did not remove the effect and that the participants were oblivious to it. We call this effect the \emph{obsolete requirements effect} (or the Gren-Svensson error), which is a bias due to the presence of obsolete requirements during effort estimation.

There was no big differences in results across the experiments, which is in favour of there being a real effect. There were different levels of experience in subjects between experiments, but they are hard to use as a moderator variable in our conceptually very different replications. The comparison can instead be done between experiments and we did find an effect even with industry participant who have much more experience, on average, than students. The effect of experience (and other moderating variables) within each experiment was averaged out since we randomized the participants into groups A, B and C. The moderating variables we did not succeed in averaging out were addressed by changing design or context (like length of the requirement specification or using industry participants). One parameter to take into account in the first analysis was study year for students (second or third year at university). Adding this variable did not create a better model (see Supplementary Material). Moreover, the effect is smaller, but still exists, the closer we get to a field setting, which is no surprise (\citet{jorgensen}).

With such a large sample ($N=461)$ and inclusion of both students and practitioners from industry, there are good reason for generalizing the findings to the larger population of people working with requirements written in natural language or in form of user stories. The over-estimations investigated in \citet{aranda} could not be explained by the subjects' experience of cost estimations, and results in this study were also apparent both in an industrial setting and by using students at the university. 

By using both students and practitioners in different settings, one recommendation is that obsolete requirements should not be visible in estimation exercises, even if we do not know the details of their effect in each specific case. This has the implications that both researchers and practitioners should take the \emph{obsolete requirements effect} into account when researching\slash working with requirements and avoid or block that effect. Requirements not needed for the estimation will still affect the decision-makers assessment of, maybe, the complexity and therefore also the estimates. 
Therefore, a specific definition and a further quantification of this error could be helpful to both practitioners and researchers even if it could be seen as a special case of the anchoring effect. Specific adjustment of this systematic error would then be possible by simply adjusting the estimates systematically, which could be done by updating our prediction model with more data.

Another great advantage of using BDA is that we share, not only our raw data, but also our posterior distributions for all parameters (see Supplementary Material). Since our posteriors then can be used as prior information in further studies, one can find effects in very small random samples. BDA also implements the fact that extraordinary claims then require extraordinary evidence and we do not reset our parameters after each replication.

\subsection{Psychological explanations}\label{sec:psych}
\citet{gren2017possible} offered two different psychological explanations for their found effect in a sample of 150 students. The first one was the \emph{representativeness heuristic}, which is a mental shortcut to lessen the cognitive load. When the needed information is not available when having to make a decision, people use similar or previous experiences instead. This often works well, but not always \citep{1982juu}. The representativeness heuristic is based on two components that are used to assign a subjective probability: (1) its similarity in characteristics to the parent population and (2) its reflection of the salient features of the process by which it is generated. In the case of these experiments, the larger requirements specification in Group C would be more probable since it is then more representative of a larger system \citep{heuristics}. Although this is an interesting explanation, it is not an obvious explanation of the entire effect found.  

The second explanation given in \citet{gren2017possible} is the \emph{decoy effect}, which is more in relation to categorical choices than extra information. The axiom of independence of irrelevant choices states that extra irrelevant options visible to the decision-maker should not affect the choice, but in some contexts, it does \citep{huber}. Since the decision in this case is not about choosing between options, this explanation is quite far away from the context of this current study and we do not think it explains any aspect of the effect found. 

The more obvious choice not mentioned in \citet{gren2017possible}, is the cognitive bias called \emph{anchoring-and-adjustment}, first published in \citet{representativeness}. This bias appears when a random number (the anchor) is presented to the decision-maker before the actual task, which then systematically influences the following decision toward that number. \citet{representativeness} state that the anchor is usually numerical, which implies that it does not have to be. In the case of the experiments in this study, obsolete requirements might have anchored a larger implementation effort in the subjects and their following estimations. This would explain the whole effect found if Groups C and B were the same in Experiments 1--4, however, Group C was also significantly higher than B. Perhaps, the effect is a combination of the representativeness heuristic and the anchoring effect in that the obsolete requirements became anchors, but a software system with changes in requirements became more representative of a software project prone to change even more in the future. It is important to note that these explanations are still speculative and need deeper investigation into the mental process of the subjects. The qualitative results of this study, i.e.\ asking the participants to explain their thought-process (see Section \ref{QualRes}), showed that people exposed to obsolete requirements cannot articulate an explanation. The participants were clueless and tried to adjust their estimations based on other or no information on what caused the estimation error. Therefore, more detailed and focused psychological experiments are needed to fully understand the results of this study from a psychological perspective.

\subsection{How to avoid the pitfall of large estimation errors}\label{sec:howtoavoid}
Even if the exact mental process behind the error is still not yet known, there are quite many findings within psychology and management of how to deal with estimation error in general to mitigate its effects (see e.g.\ \citet{hoch1996psychological}) and even some studies within the software development effort estimation (see e.g.\ \citet{connolly1997decomposed}). The former suggests that a linear decision model is to be used together with a computerized database of historical data (a more statistical approach) for more accurate forecasts in general. The latter suggests a more hands-on approach to not conducting overestimation in the software development case. Such an approach includes a couple of (decomposed) estimates instead of one (holistic) estimate. Instead of connecting the given estimation task to one single other representative experience in mind, the assessor instead had to reassess the situation and present a confidence interval with at least a lower, most probable, and upper bound. This significantly gave better estimates according to \citet{connolly1997decomposed}. They also showed that the quality of the estimates declined with increased task difficulty \citep{connolly1997decomposed}. Other known approaches to forecasting could be group forecasting, like e.g.\ the Delphi technique \citep{rowe1999delphi}. In this study, the results show that even with group forecasting, decomposed tasks or historical data in the model, we would not avoid the Gren-Svensson error since it affects all these techniques when the estimations are conducted. Decomposing tasks would work if that implies to not show any obsolete tasks next to the decomposed ones. 

Practitioners should probably spend extra time cleaning requirements specifications and backlogs from obsolete requirements, but also always give decomposed estimates and, through this process, think more in terms of probabilities instead of only efforts. For some practical guidelines on how to do this in the context of expert-judgment-based software effort estimation, see e.g.\ \citet{jorgensen2005practical}.


\section{Threats to Validity}\label{sec:limits}
Despite our efforts in addressing the validity threats after each experiment, there are still potential threats to our study. 

In this study, the artifacts, i.e.\ the requirements to estimate, could have negatively affected the experiment and thus the outcome for the student subjects. Since the requirements used in the student experiments were real requirements from industry, the students may not be experts in the area of the requirements. However, this threat was mitigated in two ways, (1) the used requirements for the students were general requirements from a domain that the students were familiar with, and (2) the last set of experiments in this study included industry subjects with requirements from their companies and domains. Another threat could be the selection of the student participants, which may influence the results since the student subjects were not volunteers as the experiment was performed as part of their courses. However, the results of the experiment did not affect the grading in the courses.

Performing experiments with only students as subjects may be a large threat concerning the representatives when compared to industry professionals. Therefore, we also carried out experiments with industry professionals as close to their real setting as possible to mitigate this threat. Another threat concerns the number of requirements used in the experiments. The first set of experiments only included four and five requirements, which is not a realistic set of requirements. Therefore, we increased the number of requirements in several steps and in the final experiment (Experiment 6) we used the real and complete product and sprint backlogs that are used in the companies. In Experiment 6, one threat to the results could be the modification, i.e.\ to mark a set of requirements as obsolete, of the requirements in the product backlogs. This threat was mitigated by having the ``gate-keepers'' at each company to modify the requirements in exactly the same way as they always mark\slash state that a requirement is obsolete. Thus, we believe that this threat has a limited effect on the results of Experiment 6.

In order to mitigate the conclusion validity, i.e.\ the ability to draw correct conclusions, we performed statistical analysis of the gathered data. By using BDA we model uncertainty both in parameters and in our models, which increases the confidence in the result. Another threat to conclusion validity may be the number of participants in the experiments. This threat was mitigated in this study by having 461 participants in total, of which 359 were student participants and 102 industry participants. However, one threat to the results of this study is that the industrial sample is much smaller than the student sample (102 versus 359), thus more studies with subjects from industry is needed. Note here that the industry participants who estimated in teams in Experiment 6 were the same individuals as in Experiment 5. Thus, we only counted them once, even if they provided data, first as individuals and then in their teams.

\section{Conclusions and Future Work}\label{sec:concl}
This paper set out to investigate if obsolete requirements have an effect on effort estimations in software development. Through a family of six experiments with both students and practitioners, the results show that having visible obsolete requirements to an assessor results in over-estimation. Thus, this obsolete requirements effect (or the Gren-Svensson error) should be taken into account when researching or conducting effort estimation. These findings are important contributions to both research, but perhaps primarily, practice since over-estimations due to obsolete requirements could possibly be avoided.

An interesting next step for this research would be to see what kind of extra requirements increases (or decreases) the effort estimates. There are undeniably cognitive aspects that influence software effort estimation that are not taken into consideration enough. Future research includes testing the suggested decomposed estimation method presented by \citet{connolly1997decomposed} in a replication of this experiment. Then, it would be possible to see if the estimates are more accurate or, if at least, get larger variance in the decomposed estimates that could be used to trigger alarm bells for decision-makers in context. Further replications that make use of our posterior distributions and help quantify the found effect further are needed. 

Finally, it would be interesting to analyze if the same effect appears in the context of a tender or down-bidding since the contexts of the estimation in this current study did not have that kind of cost-cutting demand.

\bibliographystyle{spbasic}      
\bibliography{refs}   

\begin{thebibliography}{31}
\providecommand{\natexlab}[1]{#1}
\providecommand{\url}[1]{{#1}}
\providecommand{\urlprefix}{URL }
\expandafter\ifx\csname urlstyle\endcsname\relax
  \providecommand{\doi}[1]{DOI~\discretionary{}{}{}#1}\else
  \providecommand{\doi}{DOI~\discretionary{}{}{}\begingroup
  \urlstyle{rm}\Url}\fi
\providecommand{\eprint}[2][]{\url{#2}}

\bibitem[{Aranda and Easterbrook(2005)}]{aranda}
Aranda J, Easterbrook S (2005) Anchoring and adjustment in software estimation.
  In: Proceedings of the 10th European Software Engineering Conference Held
  Jointly with 13th ACM SIGSOFT International Symposium on Foundations of
  Software Engineering, ACM, New York, NY, USA, ESEC/FSE—-13, pp 346--355,
  \doi{10.1145/1081706.1081761}

\bibitem[{Baldassarre et~al(2014)Baldassarre, Carver, Dieste~Tubio, and
  Juristo~Juzgado}]{baldassarre2014replication}
Baldassarre MT, Carver J, Dieste~Tubio O, Juristo~Juzgado N (2014) Replication
  types: Towards a shared taxonomy. In: Proceedings of the 18th international
  conference on evaluation and assessment in software engineering, ACM, p~4

\bibitem[{Bernardo(1975)}]{bernardo1975non}
Bernardo JM (1975) Non-informative prior distributions: {A} subjectivist
  approach. Bulletin of the International Statistical Institute 46:94--97

\bibitem[{B{\"u}rkner et~al(2017)}]{burkner2017brms}
B{\"u}rkner PC, et~al (2017) brms: An r package for bayesian multilevel models
  using stan. Journal of Statistical Software 80(1):1--28

\bibitem[{Cohen(1992)}]{cohen}
Cohen J (1992) Quantitative methods in psychology - a power primer.
  Psychological Bulletin 112(1):155--159

\bibitem[{Connolly and Dean(1997)}]{connolly1997decomposed}
Connolly T, Dean D (1997) Decomposed versus holistic estimates of effort
  required for software writing tasks. Management Science 43(7):1029--1045

\bibitem[{Furia et~al(2019)Furia, Feldt, and Torkar}]{furia2019bayesian}
Furia CA, Feldt R, Torkar R (2019) Bayesian data analysis in empirical software
  engineering research. IEEE Transactions on Software Engineering

\bibitem[{Gelman et~al(2008)Gelman, Jakulin, Pittau, Su
  et~al}]{gelman2008weakly}
Gelman A, Jakulin A, Pittau MG, Su YS, et~al (2008) A weakly informative
  default prior distribution for logistic and other regression models. The
  Annals of Applied Statistics 2(4):1360--1383

\bibitem[{Gelman et~al(2019)Gelman, Goodrich, Gabry, and Vehtari}]{gelman2019r}
Gelman A, Goodrich B, Gabry J, Vehtari A (2019) R-squared for bayesian
  regression models. The American Statistician pp 1--7

\bibitem[{Gren et~al(2017)Gren, Svensson, and
  Unterkalmsteiner}]{gren2017possible}
Gren L, Svensson RB, Unterkalmsteiner M (2017) Is it possible to disregard
  obsolete requirements?: an initial experiment on a potentially new bias in
  software effort estimation. In: Proceedings of the 10th International
  Workshop on Cooperative and Human Aspects of Software Engineering, IEEE
  Press, pp 56--61

\bibitem[{Halkjelsvik and J{\o}rgensen(2012)}]{halkjelsvik2012from}
Halkjelsvik T, J{\o}rgensen M (2012) From origami to software development: {A}
  review of studies on judgment-based predictions of performance time.
  Psychological Bulletin 138(2):238--271

\bibitem[{Hoch and Schkade(1996)}]{hoch1996psychological}
Hoch SJ, Schkade DA (1996) A psychological approach to decision support
  systems. Management Science 42(1):51--64

\bibitem[{Huber et~al(1982)Huber, Payne, and Puto}]{huber}
Huber J, Payne JW, Puto C (1982) Adding asymmetrically dominated alternatives:
  Violations of regularity and the similarity hypothesis. Journal of consumer
  research pp 90--98

\bibitem[{J{\o}rgensen(2005)}]{jorgensen2005practical}
J{\o}rgensen M (2005) Practical guidelines for expert-judgment-based software
  effort estimation. Software, IEEE 22(3):57--63

\bibitem[{J{\o}rgensen and Grimstad(2011)}]{jorgensen}
J{\o}rgensen M, Grimstad S (2011) The impact of irrelevant and misleading
  information on software development effort estimates: A randomized controlled
  field experiment. IEEE Transactions on Software Engineering 37(5):695--707,
  \doi{10.1109/TSE.2010.78}

\bibitem[{J{\o}rgensen and Sj{\o}berg(2001)}]{jorgen2}
J{\o}rgensen M, Sj{\o}berg DI (2001) Impact of effort estimates on software
  project work. Information and Software Technology 43(15):939--948,
  \doi{http://dx.doi.org/10.1016/S0950-5849(01)00203-8}

\bibitem[{J{\o}rgensen and Sj{\o}berg(2004)}]{jorgen3}
J{\o}rgensen M, Sj{\o}berg DI (2004) The impact of customer expectation on
  software development effort estimates. International Journal of Project
  Management 22(4):317--325,
  \doi{http://dx.doi.org/10.1016/S0263-7863(03)00085-1}

\bibitem[{Kahneman and Tversky(1974)}]{representativeness}
Kahneman D, Tversky A (1974) Subjective probability: A judgment of
  representativeness. In: The Concept of Probability in Psychological
  Experiments, Springer, pp 25--48

\bibitem[{Kahneman et~al(1982)Kahneman, Slovic, and Tversky}]{1982juu}
Kahneman D, Slovic P, Tversky A (1982) Judgement under uncertainty: Heuristics
  and biases. Cambridge U.P., Cambridge

\bibitem[{Landeta(2006)}]{delphi}
Landeta J (2006) Current validity of the delphi method in social sciences.
  Technological forecasting and social change 73(5):467--482

\bibitem[{McElreath(2016)}]{McElreath2016sra}
McElreath R (2016) Statistical rethinking: {A} {B}ayesian course with examples
  in {R} and {S}tan. CRC Press Taylor {\&} Francis Group, Boca Raton

\bibitem[{McShane et~al(2019)McShane, Gal, Gelman, Robert, and
  Tackett}]{mcshane2019abandon}
McShane BB, Gal D, Gelman A, Robert C, Tackett JL (2019) Abandon statistical
  significance. The American Statistician 73(sup1):235--245

\bibitem[{Munaf{\`o} et~al(2017)Munaf{\`o}, Nosek, Bishop, Button, Chambers,
  Du~Sert, Simonsohn, Wagenmakers, Ware, and Ioannidis}]{munafo2017manifesto}
Munaf{\`o} MR, Nosek BA, Bishop DV, Button KS, Chambers CD, Du~Sert NP,
  Simonsohn U, Wagenmakers EJ, Ware JJ, Ioannidis JP (2017) A manifesto for
  reproducible science. Nature human behaviour 1(1):0021

\bibitem[{de~Oliveira~Neto et~al(2019)de~Oliveira~Neto, Torkar, Feldt, Gren,
  Furia, and Huang}]{de2019evolution}
de~Oliveira~Neto FG, Torkar R, Feldt R, Gren L, Furia CA, Huang Z (2019)
  Evolution of statistical analysis in empirical software engineering research:
  Current state and steps forward. Journal of Systems and Software 156:246--267

\bibitem[{Rowe and Wright(1999)}]{rowe1999delphi}
Rowe G, Wright G (1999) The delphi technique as a forecasting tool: issues and
  analysis. International journal of forecasting 15(4):353--375

\bibitem[{Van~de Schoot et~al(2014)Van~de Schoot, Kaplan, Denissen, Asendorpf,
  Neyer, and van Aken}]{van2014gentle}
Van~de Schoot R, Kaplan D, Denissen J, Asendorpf JB, Neyer FJ, van Aken MA
  (2014) A gentle introduction to {B}ayesian analysis: {A}pplications to
  developmental research. Child development 85(3):842--860

\bibitem[{Stefan et~al(2019)Stefan, Evans, and
  Wagenmakers}]{stefan2019practical}
Stefan A, Evans NJ, Wagenmakers EJ (2019) Practical challenges and
  methodological flexibility in prior elicitation

\bibitem[{Tversky and Kahneman(1974)}]{heuristics}
Tversky A, Kahneman D (1974) Judgment under uncertainty: Heuristics and biases.
  Science 185(4157):1124--1131

\bibitem[{Wilson and Collins(2019)}]{wilson2019ten}
Wilson RC, Collins AG (2019) Ten simple rules for the computational modeling of
  behavioral data. eLife 8:e49547

\bibitem[{Wnuk et~al(2013)Wnuk, Gorschek, and Zahda}]{wnuk2013obsolete}
Wnuk K, Gorschek T, Zahda S (2013) Obsolete software requirements. Information
  and Software Technology 55(6):921--940

\bibitem[{Zhang and Zhang(2007)}]{zang}
Zhang T, Zhang D (2007) Agent-based simulation of consumer purchase
  decision-making and the decoy effect. Journal of business research
  60(8):912--922

\end{thebibliography}

\end{document}